\documentclass[oldversion]{aa} 
\usepackage{graphicx}
\usepackage{aalongtable}
\usepackage{txfonts}
\usepackage{rotating}
\usepackage{lscape}
\newcommand{\gsim}{\;\lower.6ex\hbox{$\sim$}\kern-7.75pt\raise.65ex\hbox{$>$}\;}
\newcommand{\lsim}{\;\lower.6ex\hbox{$\sim$}\kern-7.75pt\raise.65ex\hbox{$<$}\;}

\begin{document}
\title{Detailed abundances of a large sample of giant stars in M~54 and in the
Sagittarius nucleus\thanks{Based on observations collected at 
ESO telescopes under programme 081.D-286}\fnmsep\thanks{
   Tables 2, 3, 5, 6, and 7 are only available in electronic form at the CDS via anonymous
   ftp to {\tt cdsarc.u-strasbg.fr} (130.79.128.5) or via
   {\tt http://cdsweb.u-strasbg.fr/cgi-bin/qcat?J/A+A/???/???}}
 }

\author{
E. Carretta\inst{1},
A. Bragaglia\inst{1},
R.G. Gratton\inst{2},
S. Lucatello\inst{2,3},
M. Bellazzini\inst{1},
G. Catanzaro\inst{4},
F. Leone\inst{5},
Y. Momany\inst{2,6},
G. Piotto\inst{7}
\and
V. D'Orazi\inst{2}.
}

\authorrunning{E. Carretta et al.}
\titlerunning{Abundance analysis in M~54}

\offprints{E. Carretta, eugenio.carretta@oabo.inaf.it}

\institute{
INAF-Osservatorio Astronomico di Bologna, Via Ranzani 1, I-40127
 Bologna, Italy
\and
INAF-Osservatorio Astronomico di Padova, Vicolo dell'Osservatorio 5, I-35122
 Padova, Italy
\and
Excellence Cluster Universe, Technische Universit\"at M\"unchen, 
 Boltzmannstr. 2, D-85748, Garching, Germany 
\and
INAF-Osservatorio Astrofisico di Catania, Via S.Sofia 78, I-95123 Catania, Italy
\and
Dipartimento di Fisica e Astronomia, Universit\`a di Catania, Via S.Sofia 78, I-95123 
Catania, Italy
\and
European Southern Observatory, Alonso de Cordova 3107, Vitacura, Santiago, Chile
\and
Dipartimento di Astronomia, Universit\`a di Padova, Vicolo dell'Osservatorio 2,
I-35122 Padova, Italy
  }

\date{}

\abstract{Homogeneous abundances of light elements, $\alpha-$elements, and
Fe-group elements from high-resolution FLAMES spectra are presented
for 76 red giant stars in NGC~6715 (M~54), a massive globular cluster (GC)
lying in the nucleus of the Sagittarius dwarf galaxy. We also derived detailed
abundances for 27 red giants belonging to the Sgr nucleus. Our abundances 
assess the intrinsic metallicity dispersion ($\sim 0.19$ dex, rms scatter) of
M~54, with the bulk of stars peaking at [Fe/H]$\sim -1.6$ and a long tail
extending to higher metallicities, similar to $\omega$ Cen. The spread in 
these probable nuclear star clusters exceeds those of most GCs: these massive
clusters are located in a region intermediate between normal GCs and dwarf
galaxies. M~54 shows the Na-O anticorrelation, typical signature of GCs,
which is instead absent in the Sgr nucleus. The light elements (Mg, Al, Si) 
participating to the high temperature
Mg-Al cycle show that the entire pattern of (anti)correlations produced by
proton-capture reactions in H-burning is clearly different between the most
metal-rich and most metal-poor components in the two most massive GCs in the
Galaxy, confirming early result based on the Na-O anticorrelation. As in
$\omega$ Cen, stars affected by most extreme processing, i.e. showing the 
signature of more massive polluters, are those of the metal-rich component.
These observations can be understood if the burst of star formation giving birth
to the metal-rich component was delayed by as much as 10-30~Myr with respect to
the metal-poor one. The evolution of these massive GCs can be easily reconciled
in the general scenario for the formation of GCs recently sketched in Carretta
et al.(2010a) taking into account that $\omega$~Cen could have already 
incorporated the
surrounding nucleus of its progenitor and lost the rest of the hosting galaxy
while the two are still observable as distinct components in M~54 and the
surrounding field.
  }

\maketitle

\section{Introduction}

NGC~6715 (M~54) is a massive globular cluster (GC) immersed in the nucleus of
the Sgr (SgrN), a dwarf galaxy currently disrupting within our Galaxy (see Ibata et
al. 1994; Bellazzini et al. 2008, from now B08, and references therein).
Therefore, in this Chinese box  game, it is at the same time the nearest 
{\it bona fide} cluster of extragalactic origin and the second most massive GC in the
Milky Way, being $\omega$~Cen the most massive one (see Harris 1996).

The complexity of GCs nature has been extensively investigated (see Gratton et al. 2004
and Charbonnel 2005 for comprehensive and recent reviews): their
stars do not satisfy the very definition of simple steller population
(SSP). On the contrary, all the evidence suggests that GCs had a
significant and very peculiar chemical evolution including self-enrichment of
metals from ejecta of massive stars in the very early phases. In turn, this is
likely related to their formation mechanisms (e.g. Carretta 2006) and to the
environment where they  formed (Carretta et al. 2010a).

On top of earlier studies (see Kraft 1994 for references), our recently
completed survey of 19 GCs (Carretta et al. 2009a, 2009b) confirmed 
star-to-star variations in light elements (O, Na, Mg, Al, Si as well as F and 
lighter elements Li, C, N, see Smith \& Martell 2003, Smith et al. 2005) in all
studied clusters. They are present over all the typical mass range spanned by
galactic GCs, from the small (mainly disk) objects like M~4 and M~71 up to the
more massive (mainly inner halo) clusters like NGC~7078 (M~15) and NGC~2808.
This universality of chemical pattern is the result of the superposition of at
least two stellar generations; they are slightly spaced in time, and distinct by
the signature of an early pollution by the ejecta that only more massive stars,
now extinct, may have been able to produce (Gratton et al. 2001) in
proton-capture reactions of H-burning at high temperature (Denisenkov \&
Denisenkova 1989; Langer et al. 1993).

However, for most of GCs studied, these observed variations in light elements
are not accompanied by a dispersion in the abundances of heavier elements
mostly produced in supernovae explosions. In fact, as far as Fe is
concerned, most GCs can still be considered mono-metallic. The upper limit to
the scatter in [Fe/H]\footnote{We adopt the usual spectroscopic notation,
$i.e.$  for any given species X [X]= log(X)$_{\rm star} -$ log(X)$_\odot$
and  log $\epsilon$(X) = log (N$_{\rm X}$/N$_{\rm H}$) + 12.0 for absolute
number density abundances.} is less than 0.05 dex, meaning that the degree of
homogeneity is better than 12\% in most cases (Carretta et al. 2009c).

On the other hand, an {\it intrinsic} spread in Fe abundances has been known 
for a long time to exist in some GCs, the most noticeable examples being M~54 and
$\omega$~Cen (see e.g. the references in Carretta et al. 2010b).
For inference, these observations call for quite large and repeated bursts of 
star formation, involving all the available range in masses, due to the
nucleosynthetic origin of heavy elements in explosive burning. Possibly also
some delay between individual bursts is required, to allow the replenishing of a 
gas pool up to some critical threshold.

Multiple (even relatively recent) episodes of star formation in these objects
carry a strong resemblance to what is found for dwarf galaxies in the local
universe (e.g. Tolstoy et al. 2009). In some cases (e.g. $\omega$~Cen, Bekki and
Freeman 2003) it has been explicitly suggested that the GCs were the nuclei of
currently disrupted dwarf galaxies (see Boker 2008 and references therein).

\begin{figure}
\centering
\includegraphics[scale=0.40]{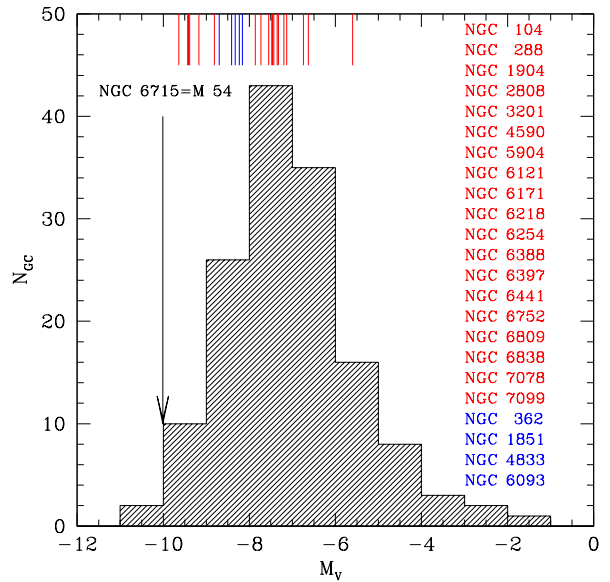}
\caption{The distribution of the total visual absolute magnitudes (a proxy for 
total mass) for all globular clusters
in the Harris (1996) catalogue. Red lines indicate the location of the GCs already
analyzed in our project ``Na-O anticorrelation and HB". Blue lines show the 
additional clusters under scrutiny and the black arrow indicates the position
of NGC~6715 (M~54) in this distribution. The most massive cluster (not indicated)
is $\omega$ Cen.}
\label{f:histomv}
\end{figure}

The nearby, extragalactic cluster M~54 is a key object, being the  most massive
of the four GCs associated to the merging Sgr dwarf spheroidal (dSph) galaxy and
the second most massive of all MW GCs, as shown in Fig.~\ref{f:histomv}. M~54 is
old and metal-poor (e.g. Layden and Sarajedini 1997 found [Fe/H$=-1.79$  and the
same age of GCs of similar metallicity).
By obtaining and studying the full extent of the Na-O anticorrelation,
the most striking signal of early pollution in GCs, 
we can answer key questions about the still obscure formation process of
these old stellar aggregates: (i) did GCs form through the same path in
every galactic environment? (ii) is the occurrence of multiple stellar
generations, as indicated by different degrees of O-depletion/Na-enhancement,
unavoidable in the formation and first evolution of GCs in all galaxies? 

While some studies of the chemical composition of red giant branch (RGB) stars in
this GC exist in the literature (Brown et al. 1999, hereinafter BWG99), they
are limited to just a few stars and hence useful only to provide some interesting
average values. What is still lacking is an extensive survey, and adequate
analysis and sampling of the chief signature of early chemical evolution in GCs.
With a collection of high resolution spectra for a large sample of cluster
stars, we can shed light on the complex scene where this initial evolution
occurred, possibly in the core of giant clouds/associations or even, in the
case of the most massive clusters like M~54, of dwarf galaxies (Bekki et al.
2007).

In Carretta et al. (2010b) we pointed out how the similarities found in the
chemistry for the two most massive clusters in the Galaxy, M~54 and $\omega$
Cen, can be simply explained assuming that they are essentially similar objects (namely,
nuclear  star clusters at the centre of dwarf galaxies). Their differences are
indeed accounted for if these GCs are observed at different stages of their
evolution. 

In this paper we present the full analysis of the data and the detailed 
chemical composition of a large sample of stars in M~54 and SgrN, providing
further evidence in support of the above scenario sketched in Carretta et al.
(2010a).

The paper is organized as follows: Section 2 describes the target selection criteria 
and observations; Section 3 outlines the analysis procedure, illustrating
also the derivation of the adopted atmospheric parameters and their uncertainties.
Section 4 deals with the Fe distribution, while in Section 5 and 6 the Na-O and Mg-Al 
anticorrelations are respectively discussed. $\alpha$ and Fe-peak elements 
abundances in the program stars are described in Section 7; Section 8 
examines our current understanding of GC formation scenario and, 
finally, a summary and conclusions are given in Section 9.

\section{Target selection, observations and input data}

Photometric data ($V,I$ magnitudes) were taken from Monaco et al. (2002), in
addition to unpublished $B$ magnitudes provided by one of the authors (MB).

\paragraph{Selection of targets and observations} This step is not trivial in
the complex  M~54$+$Sgr field, where the cluster and nuclear components are
intermingled (see e.g. Siegel et al. 2007, B08, Giuffrida et al. 
2009). Moreover, since the RGB of M~54 shows a colour dispersion, a  particular
care is needed to select genuine first-ascent red giants and not  asymptotic
giant branch (AGB) stars.

For  this task  we relied  on a pre-release of  the ACS@HST catalog  (later
presented in Siegel  et al 2007).   The majority  of the selected
sample of RGB stars in M~54 is concentrated in  the inner $1\farcm5$,  and the 
M~54 RGB  is almost
un-recognizable   at   $\sim~2\farcm0$   from  the   cluster  
center. Nevertheless,  the final  M~54  RGB sample  included  167 stars  within
$\sim9^{\prime}$, having  $15\le~V\le16.5$. On the other  hand, the SgrN RGB
sample  included 60 stars within $\sim9^{\prime}$  and in the same $V$ range.

\begin{figure}
\centering
\includegraphics[scale=0.40]{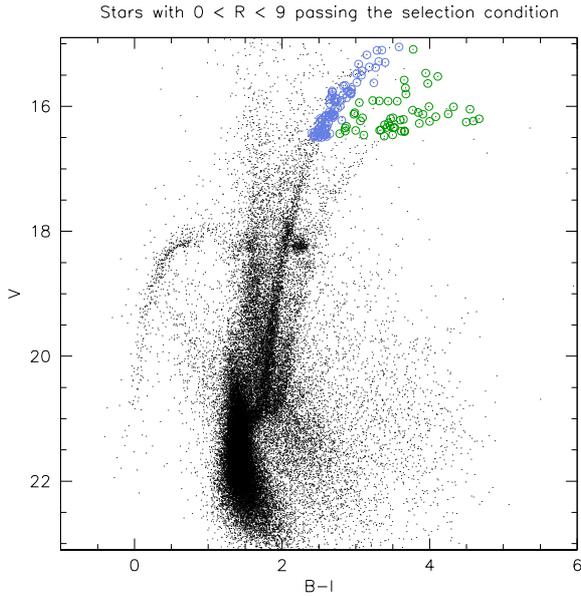}
\caption{The colour-magnitude diagram of M~54 and Sgr nucleus showing our final 
selection of targets for FLAMES observations. Green circles enclose 52 candidate
RGB stars of  Sgr and blue circles enclose 82 candidate red giant branch targets
of M~54. All  targets lie within a radial distance of 9 arcmin from the cluster
center  and do satisfy the selection criteria as described in the text.}
\label{f:selezyazan}
\end{figure}

Afterward, both samples were ``cleaned" from stars having any close neighbour
within    $\sim3\farcs0$   and    magnitudes    brighter   than    the $V+1.0$,
where $V$ is the target magnitude. Thus, the final selection included about
190 RGB stars in M~54 and Sgr; from this sample we selected the stars actually
assigned to FLAMES fibers and observed in our pointings, as shown in 
Fig.~\ref{f:selezyazan}.

The spectroscopic data were collected in service mode using the ESO high
resolution multifibre spectrograph FLAMES (Pasquini et al. 2002) mounted on the
VLT UT2. Observations were done with two GIRAFFE setups: the high-resolution
gratings HR11 (centered at 5728~\AA) and HR13 (centered at 6273~\AA),
chosen to measure the Na doublets at 5682-5688~\AA\ and
6154-6160~\AA, the [O {\sc i}] forbidden lines at 6300 and 6363~\AA, as well
as several lines of Fe-peak and $\alpha-$elements. The spectral
resolutions are R=24,200 (HR11) and R=22,500 (HR13) at the centre of
the spectra. 

We made two pointings: the first one (HR13 only) was devoted to 
observations of the metal-rich sequence of the Sgr nucleus; in this case, metal 
abundances are high enough to safely derive Na from the weak doublet at 6154-
6160~\AA. The second pointing (both HR11 and HR13) was used on the metal-
poor sequence, mostly composed by M~54 stars. Spare fibers left in the first 
pointing were also positioned on faint metal-poor stars, doubling in some case 
the exposure time in HR13 for these objects.

At the end, only 2 exposures (out of the 8 planned) were executed for the first pointing,
whereas 10 exposures with HR11 and 9 with HR13 were completed for the second one. 
This is not a source of concern since target stars of the SgrN are all quite cool
and metal-rich objects, with strong lines which can reliably be measured.
Simultaneous FLAMES-UVES observations, with resolution $R\simeq 43,000$ and
covering the spectral range from 4800 to 6800\AA, were obtained in each 
pointing. The log of observations is summarized in Table~\ref{t:logobs}.

\begin{table}
\centering
\caption{Log of FLAMES observations}
\begin{tabular}{rrcrr}
\hline
Date        &   UT       &Exp.time& Grating & airmass \\
            &            &(sec)   &         &        \\
	     \hline
Aug 09 2009 &01:12:12.226 &4500 & HR13-conf.2&1.047 \\
Aug 22 2009 &02:50:54.190 &4500 & HR13-conf.2&1.051 \\
Aug 22 2009 &04:10:34.251 &4500 & HR13-conf.2&1.220 \\
Aug 24 2009 &03:01:43.118 &4500 & HR13-conf.2&1.078 \\
Aug 24 2009 &04:17:55.094 &4500 & HR13-conf.2&1.274 \\
Sep 01 2009 &01:41:29.506 &4500 & HR13-conf.2&1.022 \\
Sep 02 2009 &02:34:33.264 &4500 & HR13-conf.2&1.091 \\
Sep 15 2009 &03:26:55.988 &4500 & HR13-conf.2&1.437 \\
Sep 16 2009 &02:47:56.318 &4500 & HR13-conf.2&1.275 \\
Aug 02 2009 &01:36:28.222 &4725 & HR11-conf.2&1.051 \\
Aug 03 2009 &00:38:24.210 &4725 & HR11-conf.2&1.149 \\
Aug 07 2009 &23:28:33.633 &4725 & HR11-conf.2&1.311 \\
Aug 08 2009 &23:40:29.190 &4725 & HR11-conf.2&1.250 \\
Aug 10 2009 &00:26:29.840 &4725 & HR11-conf.2&1.114 \\
Aug 10 2009 &01:47:44.668 &4725 & HR11-conf.2&1.014 \\
Aug 20 2009 &00:42:47.197 &4725 & HR11-conf.2&1.033 \\
Aug 20 2009 &02:03:48.847 &4725 & HR11-conf.2&1.009 \\
Aug 20 2009 &03:33:26.476 &4725 & HR11-conf.2&1.106 \\
Sep 16 2009 &01:16:06.755 &4725 & HR11-conf.2&1.056 \\
Jun 07 2009 &05:13:25.230 &4500 & HR13-conf.1&1.055 \\
Sep 19 2009 &01:41:56.426 &4500 & HR13-conf.1&1.119 \\
\hline
\end{tabular}
\label{t:logobs}
\end{table}

\paragraph{Radial velocities and membership}
We used the 1-D, wavelength calibrated spectra as reduced by the dedicated ESO
Giraffe pipeline. 
Radial velocities (RVs) were  measured
using the {\sc IRAF}\footnote{IRAF is  distributed by the National Optical
Astronomical Observatory, which are operated by the Association of Universities
for Research in Astronomy, under contract with the National Science Foundation }
task {\sc FXCORR} with appropriate templates and are shown in
Table~\ref{t:coo} (only available on line at CDS). 

Obvious interlopers in M~54 and in the nucleus of Sgr were identified 
on the basis of their RVs and disregarded 
from further analysis if the measured RV differed by more than
3$\sigma$ from the average of the cluster/nucleus stars. The mean
values for RV we derived are 143.7 ($\sigma=8.3$) and 142.4 ($\sigma=12.5$)
km/sec from the 76 stars in M54 and the 27 in SgrN, respectively, used in
the analysis. As extensively discussed in B08, the RV alone is not enough
to discriminate between M~54 and SgrN members. 

In total, we observed 76 RGB stars belonging to M~54: 7 with UVES spectra and
69 with GIRAFFE spectra (18 with only HR11 and 51 with both HR11 and HR13).
We also obtained spectra for 74 stars in the nucleus of Sgr, although not all
stars were found to be suitable for a reliable abundance analysis (see below).

\paragraph{Near infrared photometry from 2MASS.} Our 
$V$ band photometry was complemented with $K$ magnitudes to obtain atmospheric
parameters as described in the next Section. $K$ band magnitudes were obtained by
cross-correlating our photometric catalogue with the Point Source Catalogue of
2MASS (Skrutskie et al. 2006). These $K$ magnitudes were transformed to the TCS 
photometric system, as used in Alonso et al. (1999). However, due to the quite 
severe crowding in the field and to its distance, reliable near infrared magnitude 
could not be derived for all target stars (see Table~\ref{t:coo}, only available
on line at CDS).

\begin{figure} 
\centering
\includegraphics[scale=0.40]{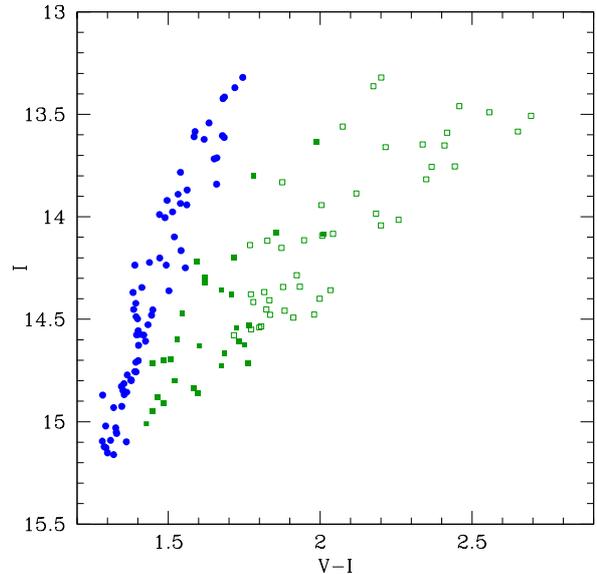}
\caption{The $I,V-I$ CMD of program stars, showing the adopted separation
between the globular cluster M~54 (blue filled circles) and the stars of the Sgr
nucleus (green squares). Filled square symbols are stars in the nucleus retained
for further analysis, whereas open squares are stars heavily affected by TiO
bands and disregarded.}
\label{f:selezm54sgr}
\end{figure}

We used the $I,~V-I$ CMD shown in Fig.~\ref{f:selezm54sgr} to separate by eye 
inspection the sequence of M~54 from the stars in SgrN. Spectra of the coolest 
stars (see Fig.~\ref{f:selezm54sgr}) in the latter component are 
heavily affected by titanium oxide (TiO) bands, which severely depress the 
continuum, in particular in the HR13 spectral range. They were omitted from 
further analysis.

After the first abundance analysis we found that three stars, attributed by us
to the SgrN on the basis of magnitudes, had a better ionization equilibrium if 
analyzed with the same relation between effective temperature and $K$ magnitude 
used for the cluster M~54 (see below). They are all very close to the cluster 
centre (within 2.5 arcmin), hence it is probable that their magnitudes are 
somewhat uncertain. Hereinafter, we assume that these stars belong to M~54.

A list of all the target stars retained in our final sample (76 RGB stars in
M~54, and 27 giant stars in Sgr nucleus) is provided in Table~\ref{t:coo} ,
together with  coordinates, magnitudes and heliocentric radial velocities 
(the table is available only in electronic form at CDS).

\section{Atmospheric parameters and analysis}

Our abundance analysis traces as closely as possible the homogeneous procedures
adopted in Carretta et al. (2006, 2007a,b,c, 2009a,b) and Gratton et al.
(2006, 2007) for measuring equivalent widths ($EW$s) and deriving atmospheric
parameters. The same set of atomic parameters and  model atmospheres were used.
Therefore, only a few points of the analysis will be summarized here.

\subsection{Atmospheric parameters}

Adopted atmospheric parameters and iron abundances are listed in
Tab.~\ref{t:atmpar} (available only in electronic form at CDS) for
all the $76+27$ program stars analyzed in the present work.

We derived first estimates of $T_{\rm eff}$ and bolometric corrections  
for all our stars, regardless of they being labelled as belonging to M~54 or SgrN, from $V-K$ 
colours. For those M~54 stars missing $K$\ magnitudes, these were obtained using 
a relation as a function of $I$\ magnitudes for stars having both filters. For 
the stars of the SgrN we derived the missing $K$ magnitudes using $V-K$ as a function 
of $V-I$. We adopted a reddening of $E(B-V)=0.14$\footnote{Monaco et al. (2005) 
concluded that no differential reddening is affecting the field from which our 
sample was selected.}, a true distance modulus of $(m-M)_0=17.10$ from Monaco et
al. (2004), and 
the relations $E(V-K) = 2.75 E(B-V)$, $A_V = 3.1 E(B-V)$, and $A_K = 0.353 E(B-V)$
from Cardelli et al. (1989). 

We refined the $T_{\rm eff}$ estimates for the M~54 stars using a relation 
between $T_{\rm eff}$ (from $V-K$ and the Alonso et al. calibration) and $K$ 
magnitudes. This approach yields much smaller internal errors than using 
colours directly (see Carretta et al. 2007a and other papers of this project), but it 
cannot be applied to the stars of SgrN because it assumes that cluster stars 
define one single sequence in the colour-magnitude diagram.

Concerning the potential contamination by AGB stars, the good agreement we
obtain, on average, between abundances based on neutral and singly ionized species
(e.g. iron) supports the atmospheric parameters we used, including surface 
gravities $\log{g}$. These were obtained from the apparent magnitudes, effective 
temperatures, distance moduli, bolometric corrections from Alonso et al. (1999), 
assuming masses of 0.85~M$_\odot$\footnote{Derived values of surface gravity 
are not very sensitive to the exact value of the adopted mass}, and $M_{\rm bol,\odot}
= 4.75$ as the bolometric magnitude for the Sun.

We derived values of the microturbulent velocities $v_t$\ by eliminating trends 
in the relation between abundances from Fe neutral lines and expected line 
strength (see Magain 1984).

Finally, we interpolated models with the proper atmospheric parameters and
whose abundances matched those derived from  Fe {\sc i} lines within the Kurucz
(1993) grid of model atmospheres (with the option for overshooting on).

\subsection{Errors in the atmospheric parameters}

The procedure for error estimates was tuned in previous papers of this project
(see Carretta et al. 2007c, 2009a,b) and will not be repeated here. We only 
point out two main differences in the error estimates with respect to the 
other analyzed GCs, both tailored to deal with an object having an {\it 
intrinsic} metallicity dispersion.

To determine the impact of errors in the adopted model metallicities on the 
final abundances, we usually considered that these errors coincide with the 
{\it r.m.s.} scatter of [Fe/H] over all stars of a GC. These estimates 
assume that the observed dispersion is an artifact due to errors.
Since we suspected a real dispersion for M~54, we used here 
the following approach. We first derived a relation expressing the 
spread in [Fe/H] for the other 19 GCs in our database (assumed to be 
homogeneous in Fe) as a bivariate function of the the median values 
of $S/N$\ and of the effective temperature of stars. We then entered 
the values appropriate for M~54 in this relation, and we obtained 0.046
dex and 0.058 dex as the values of the typical errors in [A/H] (for GIRAFFE and
UVES spectra, respectively). These values may be compared with the intrinsic
scatter of iron abundances in M~54, which is 4 times larger (see next Section).

Second, as in the other papers of this series, we considered the contribution 
of uncertainties in $EW$ to the errors in abundances as the quadratic sum of two
terms: the first one is characteristic of each line (but constant from star to
star) and can be attributed to problems in continuum tracing or blends;
the second one is characteristic of each EW measure and can be attributed
to random noise. The second term contributes to internal (i.e. star-to-star)
errors in the abundances, while the first one is important when considering
systematic abundance scale errors. To estimate correctly these terms for
M~54, we need to consider the real star-to-star dispersion in abundances.
Hence, we computed the offsets of the abundances from individual Fe~{\sc i} 
lines with respect to the mean value for each star; then for each line we 
averaged these offsets over all stars (these gives estimates for the term
characteristic of each line); and finally subtracted the average offsets from 
the abundances from individual Fe~{\sc i} lines. The quadratic average of the 
dispersions of these corrected abundances for each star represents an estimate 
of the random noise term. Applying this procedure, we derived values of 
0.077~dex and 0.060 dex (for GIRAFFE and UVES respectively). These values 
should be divided by the square root of the mean number of lines used in the 
analysis for each atomic specie.

Tables of sensitivities of abundance ratios to variations in the atmospheric 
parameters, star-to-star (internal) errors and systematic errors are given in 
the Appendix.

\section{Iron abundances and metallicity dispersion in M~54}

Homogeneity is crucial to detecting subtle effects when comparing
different clusters. Thus, adopted line lists, atomic parameters and reference
solar abundances (from Gratton et al. 2003) are strictly homogeneous for all
stars analyzed in the present program. Equivalent widths ($EW$s) were measured
with the same semi-automatic procedure described in detail in  Bragaglia et al.
(2001), with a careful  definition of the local continuum around each line, a
crucial step at the limited resolution of our GIRAFFE spectra.

The correction of $EW$ from GIRAFFE spectra to the system defined by UVES
spectra (see Carretta et al. 2009a) was deemed not necessary here, because we
found that on average the difference was negligible: 
$EW$(GIRAFFE)$-EW$(UVES)=$+0.91 \pm 0.8$ m\AA\, $rms=10.9$ m\AA\, from 209 lines
in 4 stars observed with both UVES and GIRAFFE instruments. We could then safely 
merge the results obtained with UVES and GIRAFFE.

\begin{figure} 
\centering
\includegraphics[scale=0.44]{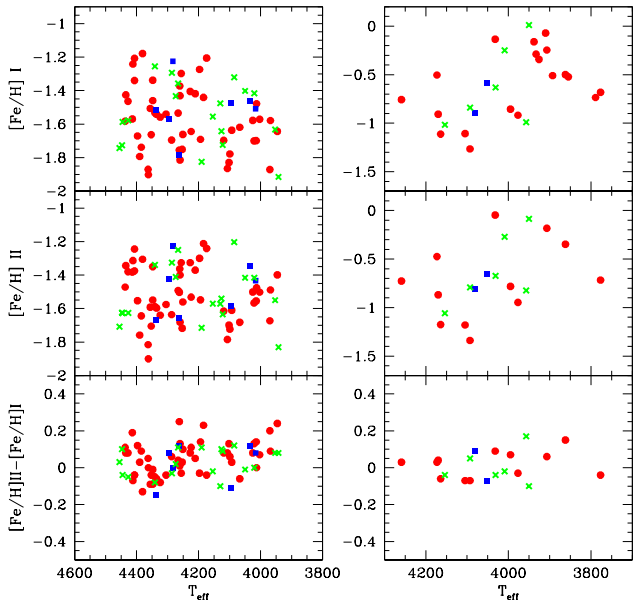}
\caption{Derived metallicity distribution of stars in M~54 (left panels:
abundances from Fe {\sc i} lines (top), from Fe {\sc ii} lines (middle) and their
difference (bottom)) and in the nucleus of Sgr (right panels) as a function of
the effective temperature. Filled red circles are stars with both HR11 and HR13
GIRAFFE spectra, green crosses stars with GIRAFFE HR13 only and blue filled
squares are stars with UVES spectra.}
\label{f:2a15e14}
\end{figure}

We plot the abundances of Fe as a function of effective temperature for the
stars belonging to M~54 and SgrN in the left and right panels of 
Fig.~\ref{f:2a15e14}, respectively. The lower panels in these Figures show good
agreement between iron abundances from neutral and singly ionized lines over
the whole range in temperature. This represents a good sanity check, since  the
ionization equilibrium for Fe is quite sensitive to any possible problem in the
abundance analysis.

The average metallicity is [Fe/H]$=-1.559 \pm 0.021$ dex ($\sigma=0.189$ dex, 76
stars) for M~54 and $=-0.622 \pm 0.068$ dex ($\sigma=0.353$ dex, 27 stars) for
SgrN, using neutral Fe lines. The average differences Fe {\sc ii} - Fe {\sc i} are
+0.041 dex ($\sigma=0.088$, 76 stars) and +0.011 dex ($\sigma=0.077$, 20 stars), 
respectively. Since the internal errors in [Fe/H] are of the order of 0.02 dex 
(average of internal errors from UVES and GIRAFFE spectra, see Appendix A) our 
first  result is that {\it the intrinsic metallicity dispersion in M~54 is confirmed 
by high resolution spectroscopy at more than 8$\sigma$} (see also Carretta et
al. 2010b).

\begin{figure} 
\centering
\includegraphics[scale=0.40]{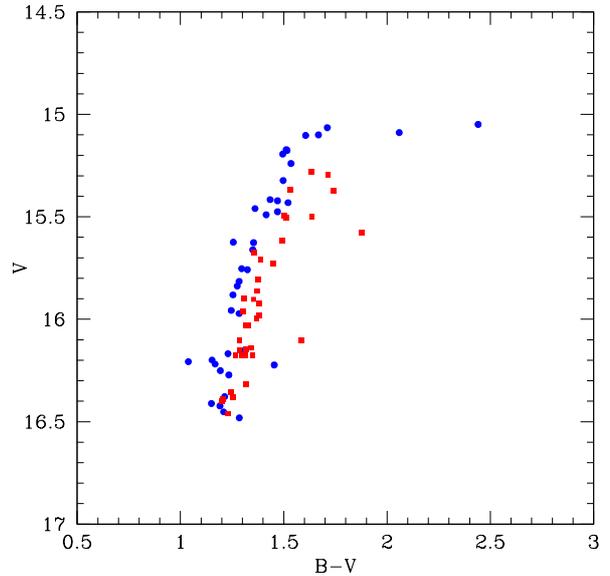}
\caption{$V,B-V$ CMD for stars in M~54. Stars with metallicity above (red filled
squares) and below (blue filled circles) the cluster average ratio [Fe/H]=-1.56
dex are marked.}
\label{f:colfe2}
\end{figure}

To show that this is not an artifact of the analysis, we marked with different
colours stars of M~54  having [Fe/H] ratios below or above the cluster average
in the $V,B-V$ CMD of Fig.~\ref{f:colfe2}. Stars with higher [Fe/H] values are 
typically redder than those with lower [Fe/H]'s along the RGB. This agrees with 
a real metallicity difference, but not with expectations from analysis errors
(stars analyzed with a temperature too low should appear as more metal-poor). 
Furthermore, we can exclude an impact of He variations: in He-rich stars the [Fe/H] ratio is
expected to be slightly higher due to the corresponding decrease in H content.
However, stars with higher He content should also be bluer than He-normal stars
(see Bragaglia et al. 2010), at odds with what observed for M~54.

\subsection{Comparison with previous results: B08}

Sixty-four stars in common with the B08 sample allow to cross-check the
reliability and the accuracy of the adopted radial velocity and metallicity
scales. It turns out that both scales display differences in the zero point at
a level less than $1\sigma$, giving no reason for concern. On the other hand, the
overall agreement is very good.

\begin{figure} 
\centering
\includegraphics[scale=0.40]{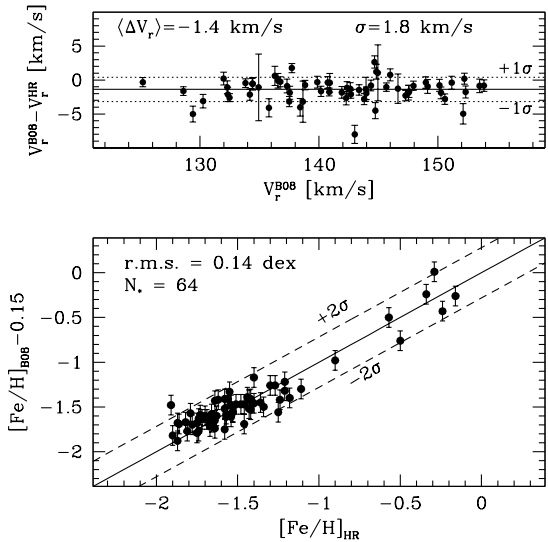}
\caption{Comparison with the RV and [Fe/H] estimates by B08.
Upper panel: differences in the radial velocity; the continuous line marks the
mean difference, the dotted lines enclose the $mean \pm 1\sigma$ range.
Lower panel: comparison between metallicity estimates (once the zero point in
B08 is taken into account); the continuous line is
the straight line of unitary slope, the dashed lines enclose 
the $\pm 2\sigma$ range about that line. The errorbars are the internal errors 
of the B08 scale (see their Fig.~5).}
\label{f:comp}
\end{figure}

The upper panel of Fig.~\ref{f:comp} shows that there is only a small 
difference in the RV zero-point between the two sets ($-1.4~km~s^{-1}$, on
average) and, most importantly, the $1\sigma$ scatter is $< 2~km~s^{-1}$, thus
confirming the results  of the tests performed by B08. We note that the
difference between the two independent estimates are smaller than $\pm 1\sigma$
in 78\% of the cases,  hence the distribution of the scatter has a peak that is
significantly narrower than a Gaussian distribution. 

A moderate difference in zero point emerges also in the comparison of the two
metallicity scales: B08 metallicities from calcium triplet are 0.15 dex higher than those
derived here, on average. The lower panel of Fig.~\ref{f:comp} shows that once
this difference is corrected for, the agreement between the two independent
sets of metallicity estimates is satisfactory over the whole 2 dex range of
metallicity considered. The $1\sigma$ scatter 
(0.14~dex) is comparable with the internal accuracy of B08 measures alone
(0.11~dex).

\subsection{Comparison with other samples: Smecker-Hane and McWilliam 2009}

There are no stars in common between our study and the samples of BWG99 (M~54 
giants) and Sbordone et al. (2007, hereinafter S07; 5 stars in Terzan~7 and 15 stars of the Sgr dSph). On the other 
hand, five of our stars are included in the sample of 14 stars analyzed by Smecker-Hane 
and McWilliam (2009, hereinafter SHM09). Their program stars were selected to 
span the full range in colour (metallicities) of RGB stars. SHM09 excluded all 
stars within 2 arcmin of the centre of M~54, and considered all their objects as 
belonging to the Sgr galaxy. The only possible exception was their star I-73, for 
which they found evidence of large processing by proton-capture reaction (see also 
McWilliam and Smecker-Hane 2005a). However, only two of the common stars are 
assigned to the SgrN according to our classification: star 38003327 is SHM09 I-150 and 
star 38004527 is SHM09 II-44, with radial distances from the centre of M~54 of 4.7 and 
9 arcmin, respectively. We assigned the other three stars, all with metallicity
[Fe/H]$< -1.4$ dex, to M~54, using the criteria
described in Sect. 2. They are star 38000597 (=SHM09 I-41) (at 3.5 arcmin from the
centre of M~54), star  38000877 (=SHM09 I-73) and star 38001507 (=SHM09 I-87) (both at
2.3 arcmin). In our 
analysis (see below), star 38000877 (=SHM09 I-73) unequivocally shows the signature 
of matter processed in H-burning at high temperature (large O-depletion, and 
Na-enhancement) typical of GC stars. On the other hand, the most metal-poor star 
in the sample by SHM09, 38000597 (=SHM09 I-41), was originally assigned also by us to 
the SgrN component, according to its colour. Only after the first abundance analysis 
it was re-classified as an M~54 star, since this provided a much better ionization 
equilibrium.

We note that almost all these stars are well within the the tidal radius of M~54 
(7.47 arcmin, Harris 1996). Admittedly (i) it is not easy to separate cluster and
nuclear components having very similar RVs and colours and (ii) a metal-poor
component is certainly present among the field stars of the Sgr dSph (e.g.
McWilliam and Smecker-Hane 2005a, Bonifacio et al. 2006, private communication).
We then caution that a clearcut separation between M~54 and SgrN is difficult
in the low metallicity regime, although it may be more probable to
find a higher fraction of M~54 stars within its tidal radius.

On average, we get [Fe/H] values lower by 0.19 dex than found by SHM09. This is
explained by the average difference in the temperatures used in the analysis:
$T_{\rm eff}(us)-T_{\rm eff}(SHM09)=-218$~K ($rms=104$~K). SHM09 adopted temperatures
derived from excitation equilibrium; they also derived photometric
temperatures, which agrees very well with ours (average offset of 6~K, with
an $r.m.s.$ of 58~K)\footnote{In our survey we prefer not to rely on temperatures
derived from the excitation equilibrium, since our sample includes also metal-poor GCs
and rather warm stars, where the limited spectral coverage of GIRAFFE HR11 and HR13 may 
not allow a large enough number of Fe lines to be reliably measured. Our photometric
approach was devised to secure more homogeneous parameters for all analyzed GCs, included 
the present one.}.

\subsection{Comparison with the metallicity distribution of other GCs}

The ideal touchstone to be compared with M~54 is obviously the most massive GC
in the Milky Way: $\omega$~Cen. A number of reasons lead to this comparison,
apart from the fact that these 2 GCs represent the high-mass tail of the GC 
mass distribution:

\begin{itemize}
\item both systems show an intrinsic dispersion in metallicity 
\item both are associated to (M~54) or suspected to be born in a dwarf galaxy
($\omega$ Cen, see Bekki and Freeman 2003)
\item both lie in the intermediate region (in the $M_V$ vs half mass radius) 
between Ultra Compact Dwarfs (UCDs) and GCs, very close to the 
low-mass limit of the UCDs (see Fig. 1 in Tolstoy et al. 2009)
\item they have a similar total mass
\end{itemize}

All these similarities lead to the legitimate suspicion that M~54 and
$\omega$ Cen might be siblings or at least next of kin.

In Carretta et al. (2010b) we  compared the metallicity distributions in
M~54$+$SgrN  (our data) and $\omega$ Cen.
Although the metallicity distribution functions (MDFs) are different, 
there are some similarities in the global appearance. Briefly:
\begin{itemize}
\item in both cases the MDF is skewed toward metal-poor stars, with the bulk 
at [Fe/H]$=-1.6 \div -1.5$ dex; 
\item this peak is followed by a gradual decrease toward higher
metallicity, up to [Fe/H]$=-1.0$ with more or less obvious secondary
peaks; 
\item finally, a long tail up to solar metallicities is observed in both 
systems, provided that stars in SgrN and in the RGB-a are included (see Carretta
et al. 2010b).
\end{itemize} 

As shown in Carretta et al. (2010b), these similarities and differences can be
easily accounted for if these are similar objects (nuclear star clusters of
dwarf galaxies), but observed at different stages of their dynamical evolution.
Within this scheme the interaction with the external environment has likely
shaped, at least partly, the dynamical and chemical properties we are
currently observing.

A generic expectation is that stellar systems closer to the Galactic centre will
be more dynamically evolved, as tidal shocks accelerate their dynamical
evolution. $\omega$ Cen is more than five times closer to the centre of our
Galaxy than the system composed by M~54$+$SgrN: dynamical evolution might
have been much faster for the former.

\begin{figure} 
\centering
\includegraphics[scale=0.44]{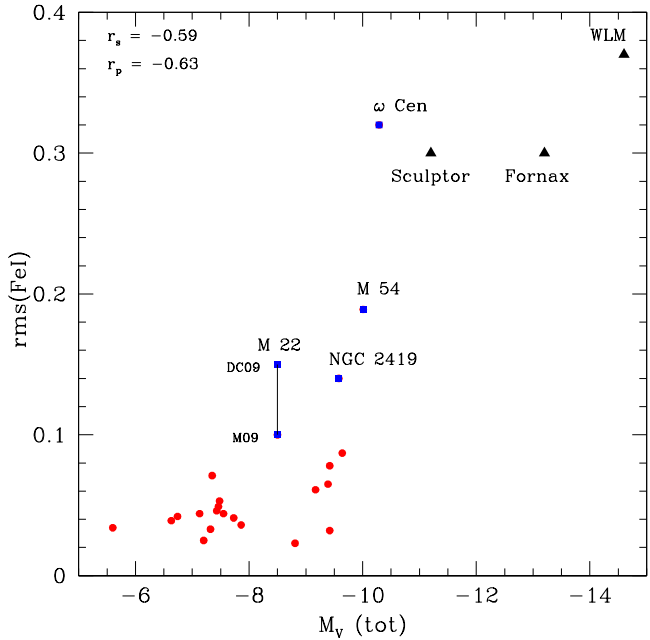}
\caption{$Rms$ scatter in Fe~{\sc i} as a function of the absolute total visual
magnitudes for the 19 GCs in Carretta et al. (2009c; red circles), for 
M~54 (present work, blue square) and for M~22, NGC~2419 and $\omega$ Cen from
the literature (blue squares). The line connects the values of the
metallicity dispersion for M~22 derived from two different studies (see text).
Black filled triangles are for the estimates of the metallicity spread in three
dwarf galaxies of the Local Group: WLM, Sculptor and Fornax. The Pearson and the
Spearman rank correlation coefficients are labeled in the figure; these values
are obtained using only the GCs.}
\label{f:rmsm15}
\end{figure}

A comparison with other Galactic GCs is provided in Fig.~\ref{f:rmsm15}, where
we show the run of the $r.m.s.$\ scatter in Fe~{\sc i} as a function of the
cluster luminosity (a proxy for the cluster mass). Red filled circles are the 19
GCs previously analyzed by us (Carretta et al. 2009c) and that can be considered
mono-metallic, as far as the iron abundance is concerned. We found that the
(small) intrinsic scatter in these GCs does correlate with cluster mass; this
evidence may be explained if GCs originating from more massive precursors are
more able to retain any inhomogeneities established in the previous phase of
rapid enrichment (see Carretta et al. 2009c, 2010a).

In Fig.~\ref{f:rmsm15} we also plot the spread in iron  derived for a few
other clusters where the analysis of individual stars was possible from
high/intermediate dispersion spectroscopy. For M~22, a line connects the values
resulting from Da Costa et al. (2009, based on a calibration of the $EW$s of the
Ca infrared triplet) and from Marino et al. (2009, from direct  high resolution
spectroscopy, more similar in concept to our approach). The other GCs considered
are M~54 (this study), $\omega$ Cen (NDC95; the  derived spread, 0.32 dex, is
exactly the same value recently obtained by  Johnson et al. (2009, hereinafter
CJ09) and the distant halo cluster NGC~2419 (Cohen 2009). 
It is not easy to assess whether the three most massive clusters in our Galaxy
define a separate sequence as a function of the total mass (luminosity).
Taken together, all clusters shown in this Figure seems to trace a relation
between intrinsic spread in iron and $M_V$ with high statistical significance,
with a steep increase at the high mass end.

Adding in this Figure also the metallicity spreads derived for three dwarf
galaxies in the Local Group (WLM, Sculptor and Fornax)\footnote{The dispersion
in [Fe/H] for WLM is directly given by Leaman et al. (2009). For Sculptor and
Fornax, we obtained the $1\sigma$ rms scatter from the FWHM of the histograms of
metallicities presented by Tolstoy et al. (2004) and Battaglia et al. (2006),
respectively.} it is interesting to note that the metallicity dispersions are
not larger by order of magnitudes, in these galaxies. The
intrinsic spreads observed in likely nuclear star clusters like M~54 and
$\omega$ Cen are comparable with those measured in small galaxies. The
latter seem to continue at higher masses the trend defined by GCs whose
most massive representatives are located in a region intermediate between the
ones of ``normal" GCs and dwarf galaxies. Actually, it would be interesting to
add to this plane also UCDs ($M_V$ from -10 to about -14.5, see Tolstoy et al.
2009); unfortunately, a simple estimate of the metallicity spread from
spectroscopic observations of individual stars is not yet available.

\begin{figure} 
\centering
\includegraphics[scale=0.40]{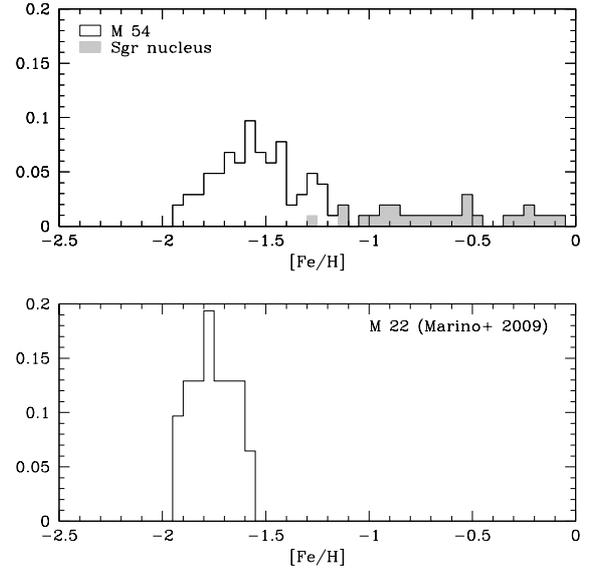}
\caption{MDF of our total sample (M~54$+$SgrN) 
as compared to M~22 from a sample of 31 giant stars by Marino 
et al. (2009). Both distributions are normalized to the total number of objects 
in each sample.}
\label{f:metM54SGRM22}
\end{figure}

Finally, while the number of stars observed in NGC~2419 is still too small, we 
note that the MDF of M~22 seems to be different from that of the two most 
massive clusters. In particular, Fig.~\ref{f:metM54SGRM22} shows that in M~22
the MDF is approximately bell-shaped, with a moderate spread around a 
metal-poor value, totally lacking the minor multiple peaks at higher 
metallicity observed both in M~54 and $\omega$ Cen (Carretta et al. 2010b).
Moreover, the long tail at almost solar metallicities is completely absent.
In the scenario described in Carretta et al. (2010a) this tail simply
corresponds to the residual component of the ancestral dwarf galaxy still
(partly) visible around the clusters. This is not surprising, because M~22
lies even closer to the central regions of the Galaxy than $\omega$ Cen, 
and it is possible that in this case the envelope of the progenitor galaxy was
completely stripped by tidal interaction with the densest Galactic
regions before the formation of more metal-rich objects.

\section{The Na-O anticorrelation in M~54}

As in previous papers (e.g. Carretta et al. 2009a and references therein), we 
used the $EW$s of the forbidden [O {\sc i}] lines at 6300.3 and 6363.8~\AA\ to
derive oxygen abundances. Abundances of Na are from $EW$s of the
two doublets of Na{\sc i} at 5682-88 and 6154-60~\AA. They were corrected for 
the effect of departures from LTE as in Gratton et al. (1999).

\begin{figure} 
\centering
\includegraphics[scale=0.40]{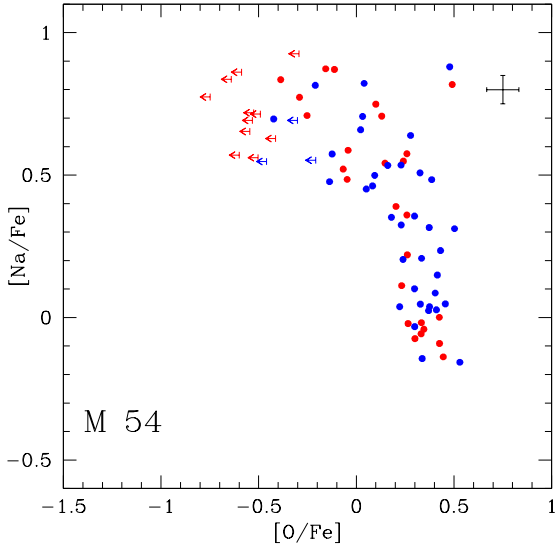}
\caption{Na-O anticorrelation in M~54 from UVES and GIRAFFE spectra; blue 
and red circles indicate respectively stars that are more metal-poor and more 
metal-rich than the cluster average [Fe/H]$=-1.56$. Upper limits in [O/Fe] are
indicated by arrows. The typical star-to-star error bars are also shown.} 
\label{f:m15antitotseq}
\end{figure}

The Na-O anticorrelation obtained for M~54 is shown in 
Fig.~\ref{f:m15antitotseq}. Despite the scatter in [Fe/H] values, metal-poor 
and metal-rich stars are well intermingled along almost the full length of the 
anticorrelation. Apart from the case of $\omega$~Cen (see below), this is the 
most pronounced known example of O-depletions anticorrelated with 
Na-enhancements among RGB stars in a GC.

The interquartile range of the distribution of [O/Na] ratios, IQR[O/Na], can be
assumed as a quantitative measure of the  extension of the anticorrelation (see
Carretta 2006); from the global sample  of 76 RGB stars in M~54 we found
IQR[O/Na]=1.169, the record in our sample of GCs. This is second in value only
to the extension we found in $\omega$ Cen (IQR[O/Na]=1.310) using data from
NDC95.

\begin{figure} 
\centering
\includegraphics[scale=0.40]{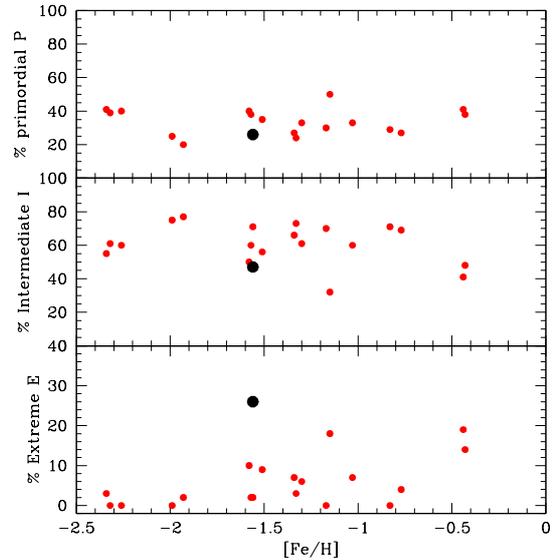}
\caption{Fraction of stars in the P, I and E  components in the 19 clusters
analyzed by Carretta et al. (2009a, red filled circles) and in M~54 (large black
circles).}
\label{f:piefe}
\end{figure}

In Carretta et al. (2009a) we separated the stars observed in each of the 19 GCs
into three distinct populations, according to their abundance ratios [O/Fe] and
[Na/Fe]. We identified (see e.g. Fig. 13 below) a primordial component P 
sharing the same Na, O
abundances of field stars with similar metallicity, and two components of second
generation stars, defining the O-poor part of the Na-O anticorrelation. These
last are separated in objects with intermediate (I) ([O/Na]$>-0.9$
dex) and extremely modified composition (E) ([O/Na]$<-0.9$ dex), with respect to
the first generation P stars.

For M~54, we found that the P fraction of first generation stars is P$_{\rm
M~54}$ = $26 \pm 6\%$, where the error is simply due to the statistics.
This value is very similar (see Fig.~\ref{f:piefe}) to the P fraction observed
in other GCs by Carretta et al. (2009a: on average P=33\% with $rms=7\%$),
regardless of their mass, metallicity and global parameters. The constant 
fraction of first generation stars is an important constrain in models of
formation of GCs.

The observed fraction of intermediate stars found in M~54 is I$_{\rm M~54}$ = 
$46 \pm 8\%$ and it does not look peculiar. However M~54 has the highest 
fraction of second generation stars with extreme composition found up to date 
in our sample of 20 GCs, exceeding the previous record detained by 
NGC~2808 (Carretta et al. 2009a): E$_{\rm M~54}$ = $28 \pm 6\%$ (see 
Fig.~\ref{f:piefe}).

\begin{figure} 
\centering
\includegraphics[scale=0.40]{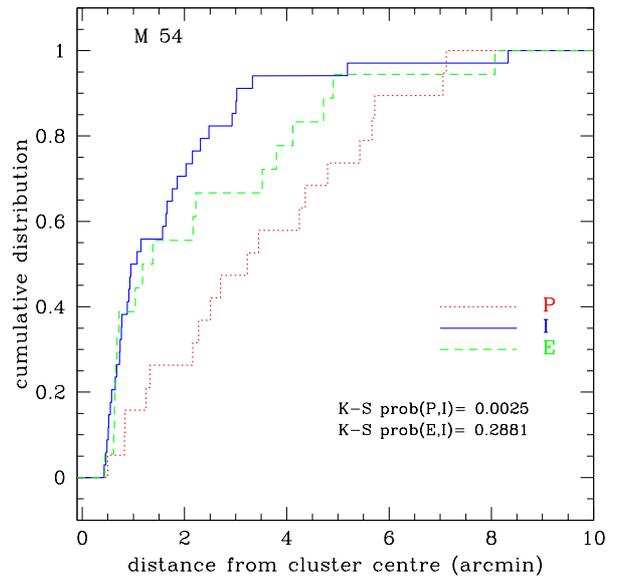}
\caption{Cumulative distributions of the distances from the cluster centre
relative to stars in M~54 belonging to the P, I, and E stellar components.}
\label{f:distrPIE}
\end{figure}

The cumulative distributions of radial distances from the centre of M~54 for the
three P, I, and E stellar components are shown in Fig.~\ref{f:distrPIE}. First 
generation stars are less concentrated than the I stars, with a quite high
level of confidence. No conclusion can be drawn with high statistical significance
about the radial distribution of the E component.

\begin{figure} 
\centering
\includegraphics[scale=0.40]{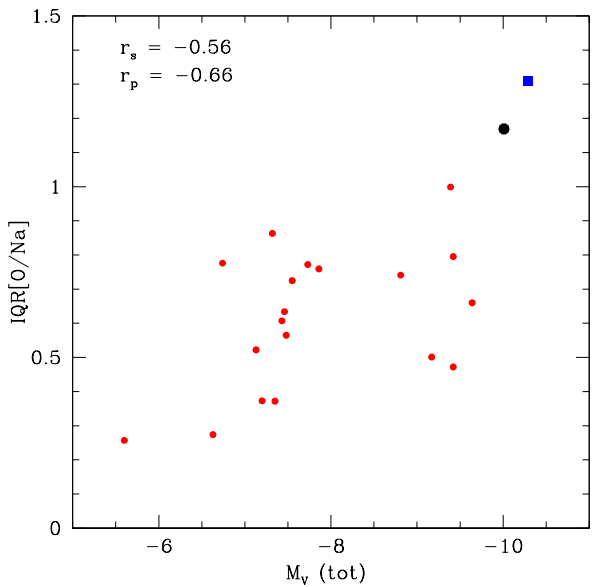}
\caption{Relation between the total absolute magnitude of clusters and the IQR[O/Na].
M~54 (circle) and $\omega$ Cen (square) are indicated by larger symbols}
\label{f:mviqr3}
\end{figure}

Summarizing, both M~54 and $\omega$ Cen (see Fig.~\ref{f:mviqr3}) nicely 
extend to the most massive GCs in the Galaxy the correlation between 
the extension of the Na-O anticorrelation and the total cluster mass (using 
the total visual absolute magnitude as proxy). This supports the definition
of GC given in Carretta et al. (2010a) as those clusters exhibiting the Na-O 
anticorrelation.

\begin{figure} 
\centering
\includegraphics[scale=0.40]{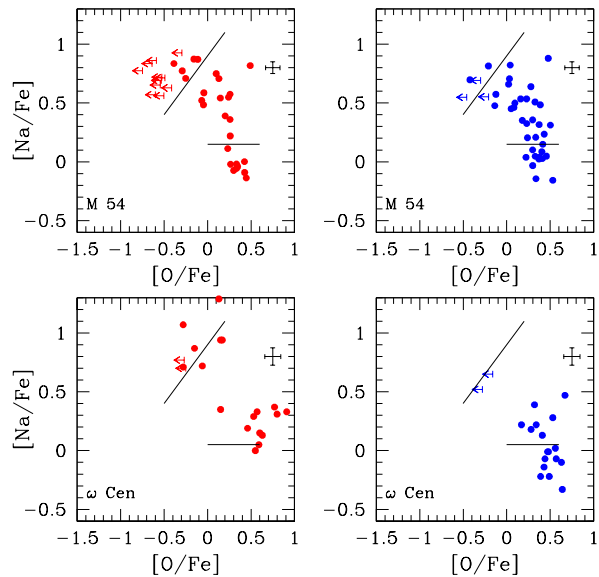}
\caption{Na-O anticorrelation in M~54 for the stars of the metal-rich (upper
left panel) and of the metal-poor (upper right panel) components analyzed here.
The corresponding lower panels show the same  in $\omega$ Cen, using the same
separation at [Fe/H]$=-1.56$ dex, from Norris and Da Costa (1995). Upper limits
in O are indicated by arrows. Lines separate the regions of the P, I and E
stellar components.}
\label{f:m15antitotseq2}
\end{figure}

\begin{figure} 
\centering
\includegraphics[scale=0.40]{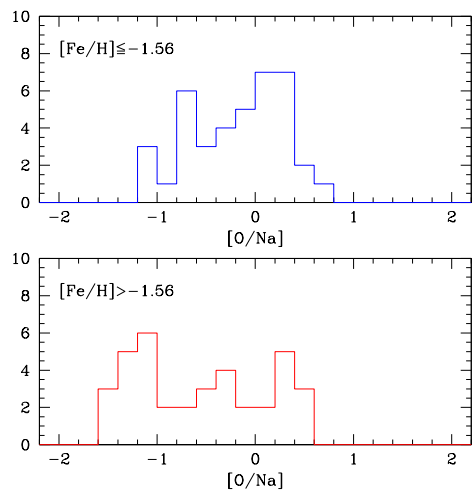}
\caption{Distribution of the [O/Na] ratios in the metal-rich (lower panel) and 
metal-poor (upper panel) components in M~54.}
\label{f:histoona}
\end{figure}

However, as anticipated in Carretta et al. (2010b), our most striking finding is 
that this feature has a clearly different extent if we consider separately the 
metal-poor and metal-rich components in M~54, as shown in the upper panels of 
Fig.~\ref{f:m15antitotseq2}. The average metallicity [Fe/H]$=-1.56$ dex is 
adopted as the separation limit in metallicity.
The minimum value [Na/Fe]$_{min}$
is the same both for the metal-rich and metal-poor population (so that the P
fractions are similar, within the errors: P$_{MP}=31 \pm 9\%$ and
P$_{MR}=24 \pm 8\%$). However, the E fractions are dramatically different:
E$_{MP}=10 \pm 5\%$ E$_{MR}=43 \pm 11\%$.
As a consequence, the extension of the Na-O anticorrelation, which is driven by
the E component, is much longer for the metal-rich component:
IQR[O/Na]$_{MR} = 1.264$, compared to a modest 
IQR[O/Na]$_{MP} = 0.783$ of the metal-poor component.
The corresponding distributions of the [O/Na] ratio, shown in
Fig.~\ref{f:histoona}, appear to be unequivocally different.

Even more interestingly, we found the same phenomenon also in
$\omega$ Cen, as shown in Fig.~\ref{f:m15antitotseq2} (lower panels; see also
Carretta et al. 2010b)! This
feature, already available in the data since the work by NDC95 seems to have 
been almost overlooked up to now\footnote{Another evidence in $\omega$ Cen also comes from
the observed anticorrelation of C and N abundances (NDC95), where the most
C-poor, N-enhanced stars belong all to the metal-rich component. However, since
C and N are also affected by standard evolutionary mixing and extra-mixing
episodes (see e.g. Charbonnel 2006, Carretta 2008, and reference therein), it is
a somewhat less clear evidence.}.
In truth, already NDC95 (p. 695, left column) noted an {\it apparent dependence
of the operation of the ON cycle on [Fe/H]}, confirming early findings by Cohen
and Bell (1986) and Paltouglu and Norris (1989). NDC95 stated that they were
{\it unable
to offer any explanation of this result for the $\omega$ Cen stars}. However, at
the time, changes in abundances resulting in the Na-O anticorrelation were
mostly interpreted as internal changes due to evolutionary mixing in the 
observed stars. Only a few years later Gratton et al. (2001) discovered the
Na-O anticorrelation among unevolved cluster stars, changing forever the
interpretative framework.

\begin{figure} 
\centering
\includegraphics[scale=0.44]{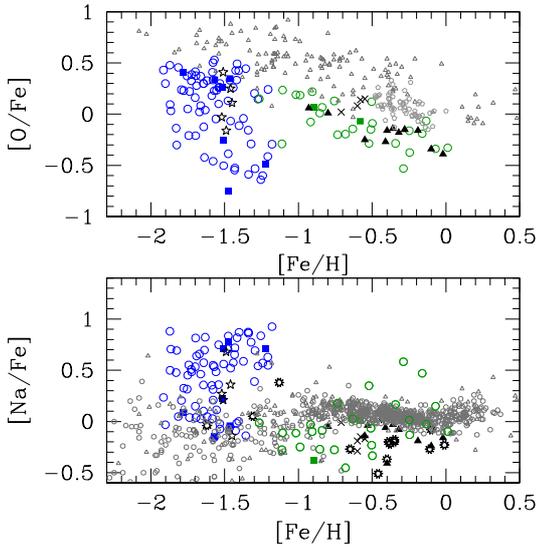}
\caption{Abundance ratios [O/Fe] (upper panel) and [Na/Fe] (lower panel) for
stars in M~54$+$Sgr in the present work and in the literature. Blue open circles
and blue filled squares are our stars in M~54 with GIRAFFE and UVES spectra, 
respectively. Green open circles and filled squares are our stars in Sgr with
GIRAFFE and UVES spectra, respectively. Open black star symbols are 5 giants in
M~54 from Brown et al. (1999); black crosses are 5 giants in Terzan 7 and
filled triangles giants in Sgr from S07. Black open ``star
explosion" symbols are Sgr giants from Smecker-Hane and McWilliam (2009, 
submitted). Finally, small grey symbols are samples of galactic field stars
(Fulbright 2000, Gratton et al. 2003, Reddy et al. 2003, Venn et al. 2004).}
\label{f:m54ona}
\end{figure}

Thus, we find that a common feature of  the two most massive GCs in the
Galaxy, both with a large metallicity spread,  is to show a different extension
of the Na-O anticorrelation in their metal-poor and metal-rich components, with
the latter reaching higher degrees of processing by proton-capture reaction in
H-burning at high temperature\footnote{This same result has been very recently
confirmed also by Marino et al. (2010) in their  ongoing analysis of a large
sample of stars in $\omega$ Cen.}.

What about the stars of the Sgr nucleus? In Fig.~\ref{f:m54ona} the run of
[O/Fe] and [Na/Fe] ratios as a function of metallicity is shown both for M~54
and SgrN from this work, and compared to literature samples: BWG99 for
M~54, S07 for Terzan~7, another GC associated to Sgr, and again S07 and SHM09
for Sgr (main body)\footnote{We do not use in this and other comparisons data
from M05 because they derived only three $\alpha-$elements (Mg, Ca, Ti) for 15
bright giants in the Sgr main body. Their abundances are nicely intermingled for
these elements to those from Sbordone et al. (2007) who studied many more atomic
species.}.

In the following, when comparing abundances in M~54 with those in the SgrN
we will consider the metal-poor and the metal-rich cluster components
(shown as blue open circles) together; the SgrN  from our study will be plotted
as green open circles. For figures referring to a single cluster (either M~54 or
$\omega$ Cen), stars in the metal-rich and metal-poor components are indicated
with filled red and blue symbols, respectively, when relevant to the discussion.

The large spread in O and Na abundances (anticorrelated with each other) is
confined only to stars in M~54. Those in SgrN present a run of these
elements as a function of the metallicity typical of Galactic field stars, 
apart from the well known offsets already observed for stars in Sgr (S07, SHM09
and references therein). The [O/Fe] ratios are clearly below the mean locus of
Galactic stars and also Na is quite deficient in Sgr stars with respect to disk
stars in the solar neighbourhood. 
The scatter we found in O abundances for SgrN stars is larger than in
the results by S07, based on higher dispersion UVES spectra. We think that 
part of our scatter is explained by our use of intermediate resolution
GIRAFFE spectra, in particular at these high metallicities, where the effect of
unrecognized blends might be more severe.
However, part of the observed scatter may also be genuine, intrinsic to the
Sgr stellar population, as shown in the bottom panel of Fig.~\ref{f:m54ona}, 
where the dispersion in Na for our data is roughly comparable with the results
by SHM09, obtained with HIRES at the Keck telescope (open ``star explosion"
symbols).

A couple of stars, attributed to SgrN on purely
photometric criteria, showed instead abundances (Fe, O,
Na) similar to those of stars in M~54. The same pattern was found by 
SHM09 in their star I-73: its
composition ([O/Fe]$=-0.91$, as quoted in McWilliam and Smecker-Hane 2005a;
[Al/Fe]$=+1.40$, [Na/Fe]$=+0.38$, [Fe/H]$=-1.13$, SHM09, with our solar
abundances) makes very likely its membership to M~54\footnote{In 2005 a
metallicity dispersion in M~54 was still not very fully assessed, however we now
have stars as metal-rich as [Fe/H]$\sim -1.2$ dex in our M~54 sample, not very
different from [Fe/H]$=-1.13$ dex of star I-73.}. Anyway, even
including these few stars in the M~54 sample our results would not change; on
the contrary, they would be strengthened, since these stars are on the
metal-rich side of the metallicity distribution in the cluster.

In summary, the GC M~54 shows a well developed, very extended Na-O
anticorrelation which is not observed in the surrounding nucleus of the Sgr
dSph. We note in passing that the same occurrence seems to hold for
another dwarf galaxy, Fornax, where evidence of stars participating to a Na-O
anticorrelation are found among its GCs (Letarte et al. 2006),
but not among field stars in the galaxy itself (Shetrone et al. 2003).
The Na-O anticorrelation is thus a distinctive signature of some event occurring
only in the formation and early evolution of GCs.
We shall return later to this point.

\section{The high temperature Mg-Al cycle}

The proton-capture reactions involving Mg, Al (and Si, see Yong et al.
2005, Carretta et al. 2009b) require a much higher temperature regime and 
consequently allow us to probe the contribution from polluters with higher mass,
on average.

Al abundances were derived from the subordinate doublet at 6696-98~\AA, only
for stars observed with UVES. The Mg abundances are typically obtained from two
to three high excitation lines, while abundances of Si (partly produced in
proton-capture reactions if the temperature is high enough) are estimated from
several transitions (see Carretta et al. 2009b for details and Gratton et al.
2003 for the atomic parameters).

\begin{figure} 
\centering
\includegraphics[scale=0.44]{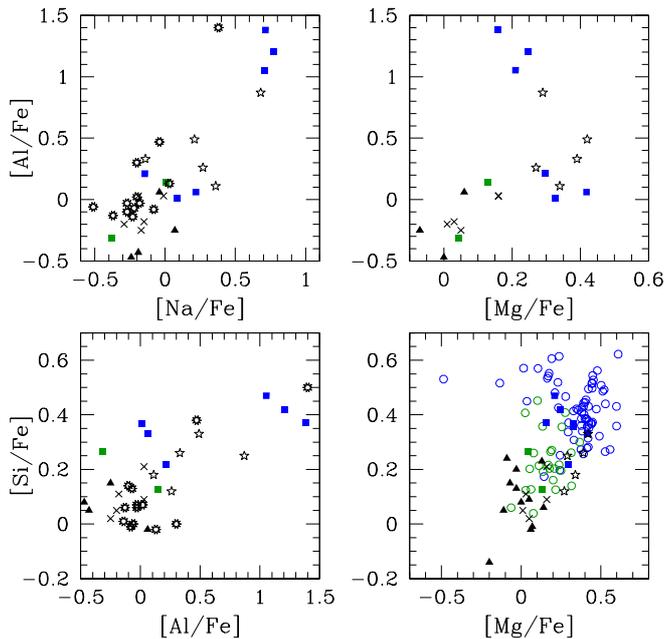}
\caption{Summary of the (anti)correlations involved in the Mg-Al cycle for stars
of SgrN and of M~54  analyzed in the present work and in
previous literature studies. Symbols are as in Fig.~\ref{f:m54ona}.}
\label{f:ciclomgal}
\end{figure}

A summary of the correlations and anticorrelations among elements involved in
the Mg-Al cycle of H burning at high temperatures is given in
Fig.~\ref{f:ciclomgal} for stars analyzed in this work as well as in previous
studies, both for M~54 and the Sgr dwarf galaxy.

A clear, strong correlation between abundances of Na and Al is seen in M~54 from
our data, nicely supported by the
five stars analyzed by BWG99. Considering all samples together, stars in SgrN 
seem to have a roughly constant Na level. 
Star SHM09=I-73 has a very high Al abundance, comparable to
some of the M~54 stars: this further supports its membership to the cluster and not 
to SgrN. In this respect, the ratio [Na/Al] can be used as a useful diagnostic. 
As discussed e.g. by CJ09 in the case of $\omega$~Cen, a constant ratio 
[Na/Al]$\sim -0.2$ dex is expected if only SNe contributed to the metal enrichment, 
because Na and Al production have a similar dependence on metallicity (Arnett 
1971, Woosley and Weaver 1995). On the other hand, in the dense environment of 
GCs, other polluters may raise the Na and Al abundances over the level possible 
to SNe alone. However, these enhancements might be (and usually are) not scaled 
in lockstep, due e.g. to the different temperature thresholds required to trigger 
the Ne-Na and Mg-Al cycles. Hence, we may regard any large deviation of the ratio 
[Na/Al] from the value of -0.2 dex as likely due to proton-capture reactions 
polluting cluster stars. This ratio is -1.02 dex for the suspected cluster star 
I-73 in SHM09, but it is -0.51 dex for the most metal-poor star ([Fe/H]$=-1.62$ dex) 
in their sample, again casting some doubts on the unambiguous classification of the 
most metal-poor objects as genuine example of the Sgr nuclear component. We note
that two other stars (I-242 and I-38) have [Na/Al]$\sim -0.5$ dex, quite
different from the value expected from SNe. However, their high metallicity 
([Fe/H]$\geq -0.46$) makes very difficult to mis-classify them as stars in M~54.

Concerning Mg and Al, a quite good anticorrelation is observed for M~54 (upper
right panel in Fig.~\ref{f:ciclomgal}), with moderate Mg depletions associated 
to large enhancements in Al abundances, both in our sample and in BWG99. This 
behaviour is evidently not present 
in the Sgr field stars, where, on the contrary, Mg and Al abundances are
possibly positively correlated with each other.

The large enhancements in Al abundances are a signature of very high temperatures
achieved during H-burning. At such high temperature we expect leakage from the 
Mg-Al chain (Yong et al. 2005, Carretta et al. 2009b) producing some fresh 
Si. This is exactly what is observed in M~54 
(Fig.~\ref{f:ciclomgal}, bottom-left panel). There seems to be a small offset in
the [Si/Fe] ratio with respect to BWG99, even after correcting for different solar
abundances. Apart from this, a good correlation is again observed only for 
cluster stars.
We also note in this panel that the two metal-poor stars in the Sgr sample of 
SHM09 (the above mentioned O-depleted star I-73 and the most metal-poor object
in their sample, star I-41, with [Fe/H]$=-1.62$ dex) are well mixed with the
stars of M~54, as also seen for the Al-Na correlation, strengthening the
conclusions reached above concerning their possible membership to M~54.

Most stars in M~54 show a correlation between the $\alpha-$elements Mg and
Si. However,
although more confused  than expected, a Mg-Si {\it anticorrelation} is rather
evident in the bottom-right panel of Fig.~\ref{f:ciclomgal}: the most Mg-poor
stars in M~54 (reaching respectable depletions of [Mg/Fe]$\sim -0.5$ dex)
are also Si-rich. On the contrary, the field stars of SgrN do show the normal
correlation expected between two elements from $\alpha-$capture processes.
  
\begin{figure} 
\centering
\includegraphics[bb=19 164 580 492, clip,scale=0.44]{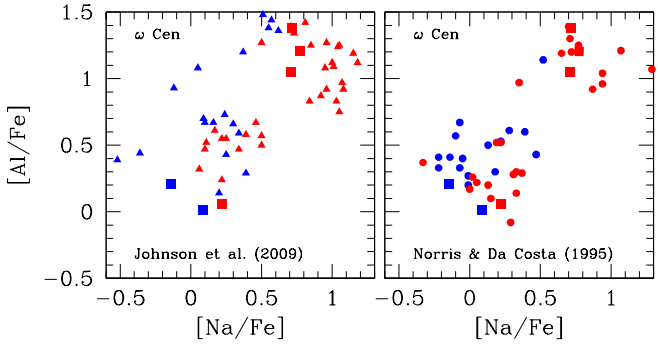}
\caption{Na-Al correlation for giants in $\omega$ Cen separated in stars more
metal-poor (blue points) and more metal-rich (red points) than [Fe/H]$=-1.56$
dex. In the left panel data are from CJ09 and in the right
panel from NDC95. The abundances are corrected to our scale of solar reference
abundances. In both panels, big filled squares are our stars in M~54 with UVES
spectra, with the same colour coding.}
\label{f:n51alna}
\end{figure}

\begin{figure} 
\centering
\includegraphics[scale=0.40]{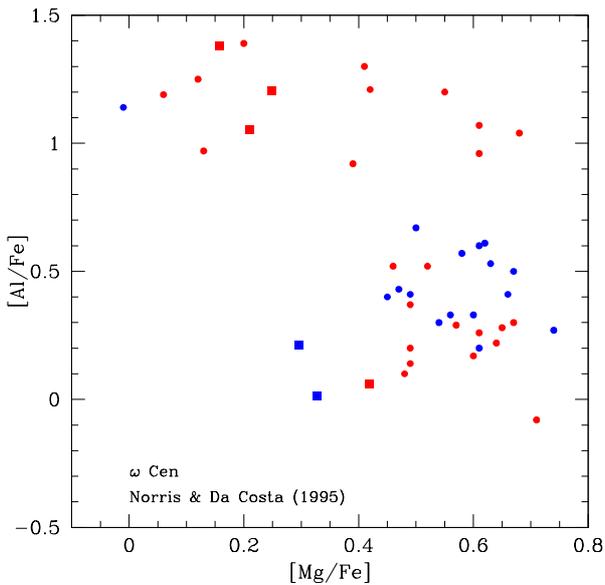}
\caption{Mg-Al anticorrelation for giants in $\omega$ Cen (from NDC95) 
separated in stars more
metal-poor (blue points) and more metal-rich (red points) than [Fe/H]$=-1.56$
dex. Big filled squares are our stars in M~54 with UVES spectra, with the
same colour coding.}
\label{f:n51almg}
\end{figure}

The observed variations in Al are quite large in M~54; this GC does not resemble
clusters like M~4, it is more NGC~2808-like. This is not unexpected since we
found (Carretta et al. 2009b) that the production of Al is a tight function of a
bi-linear combination of $M_V$ and [Fe/H], with aluminum being enhanced mostly 
in metal-poor and/or massive GCs. M~54 obviously presents both these
characteristics, thereby its stars show the signatures of the very high
temperatures reached by proton-capture reactions in H-burning  ($> 65$ MK, as
suggested by the production of $^{28}$Si, see Arnould et al. 1999) .

Unfortunately, we have only a handful of stars (the $7+2$ with UVES
spectra) where the main outcome of the Mg-Al cycle (Al) is measurable.
On the other hand, if the similarities between M~54 and $\omega$ Cen are not
merely a fortuitous coincidence, we should see the signature of polluters with
an average larger mass (temperature) also in $\omega$ Cen.
This is exactly what is shown in Fig.~\ref{f:n51alna}, Fig.~\ref{f:n51almg} 
using data from high resolution spectra by NDC95 and from intermediate
resolution spectroscopy of a large sample of giants in
$\omega$~Cen with Na and Al abundances recently derived by CJ09.
In both panels of Fig.~\ref{f:n51alna}, the Na-Al correlation has more or less
the same extent of the one we found in M~54; moreover, due to the much larger
sample, the appearance of a clear bimodality (already hinted by our limited
sample) is striking: apart from a few outliers, stars in $\omega$ Cen seem to
be neatly clustered in two distinct groups, at high-Na/high-Al and
low-Na/low-Al, respectively.

More important, in these extensive sets of abundances we retrieve, for elements
involved in the Mg-Al chain, the same phenomenon seen for elements
participating to the ON and NeNa cycles: the Na-Al correlation (as well as the
Mg-Al anticorrelation) is preferentially much more extended among cluster stars 
of the metal-rich component than in more metal-poor
ones. The same effect is also visible in our smaller sample of stars in M~54.
A metallicity dependence of the distributions of Na and Al was already noted by
CJ09: by exploiting their large sample, combined with that of
Johnson et al. (2008), they were able to trace in the Al distribution the
equivalent of the primordial, intermediate and extreme populations we have
defined from the Na-O anticorrelation. Moreover, they found that concerning the
distribution of [Al/Fe] the components of metal-rich and metal-poor stars were 
different at the 94\% level of confidence (using [Fe/H]$=-1.2$ as separating 
metallicity).

\begin{figure} 
\centering
\includegraphics[scale=0.40]{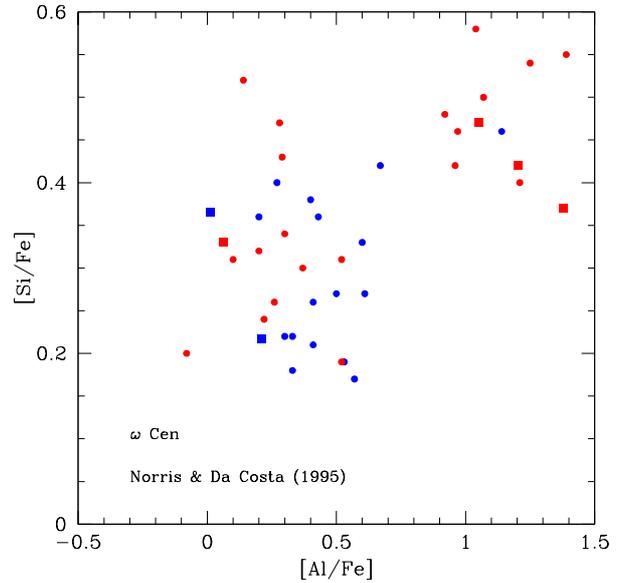}
\caption{Si-Al correlation for giants in $\omega$ Cen (from NDC95)  separated in
stars more metal-poor (blue points) and more metal-rich (red points) than
[Fe/H]$=-1.56$ dex. Big filled squares are our stars in M~54 with UVES spectra,
with the same colour coding.}
\label{f:n51sial}
\end{figure}

Finally, the same sort of segregation in metallicity can be recognized also for
the Si-Al correlation in Fig.~\ref{f:n51sial}, although less clearly; we remind
that $^{28}$Si production is the result only of a secondary {\it leakage} in the
main Mg-Al cycle.

While some hints of metallicity-dependence for light elements involved in
proton-capture reactions were already presented in the literature for $\omega$
Cen, this is the first time they are reliably assessed for M~54.
Moreover, the novelty of our findings is to consider that {\it the entire pattern of
anticorrelations and correlations established for these elements is clearly
different between the most metal-rich and most metal poor stellar components in
the two most massive clusters in the Galaxy}. We think that this cannot be 
ruled out as a mere chance coincidence and must be incorporated as a
constraint in any formation model about these objects.

\section{$\alpha-$elements and Fe-group elements}

We measured a number of lines due to several $\alpha-$ (Mg, Si, Ca and Ti) and 
Fe-peak elements (Sc, V, Cr, Mn, Co, Ni, Cu, beside obviously Fe) in our UVES and 
GIRAFFE spectra of stars of M~54 and SgrN. Mean abundances are listed in 
Table~\ref{t:meanabu}, where we provide the number of stars where lines were
measured and the $r.m.s.$ scatter of the mean. Hyperfine structure was taken 
into account for Sc, Mn, and Co following Gratton et al. (2003) and for Cu from 
Sobeck (private communication).

\setcounter{table}{3}
\begin{table*}
\centering
\caption{Mean abundances from UVES and GIRAFFE spectra for stars in M~54 and Sgr nucleus}
\begin{tabular}{lcccc}
\hline
                     &   M~54        &   M~54         &     Sgr       &   Sgr         \\
Element              &  UVES         & GIRAFFE        & UVES          & GIRAFFE       \\
                     &n~~~   avg~  $rms$ &n~~   avg~~  $rms$  &n~~   avg~~  $rms$ &n~~   avg~~  $rms$ \\        
\hline
$[$O/Fe$]${\sc i}    &7 $-$0.02 0.47 &69   +0.08 0.33 &2 $-$0.00 0.09 &25 $-$0.10 0.21\\
$[$Na/Fe$]${\sc i}   &7   +0.33 0.39 &69   +0.43 0.31 &2 $-$0.19 0.27 &25 $-$0.04 0.25\\
$[$Mg/Fe$]${\sc i}   &7   +0.28 0.09 &67   +0.35 0.18 &2   +0.09 0.06 &25   +0.16 0.11\\
$[$Al/Fe$]${\sc i}   &6   +0.65 0.62 &		      &2 $-$0.09 0.33 & 	      \\
$[$Si/Fe$]${\sc i}   &7   +0.36 0.08 &69   +0.42 0.10 &2   +0.20 0.10 &25   +0.23 0.11\\
$[$Ca/Fe$]${\sc i}   &7   +0.32 0.08 &69   +0.36 0.10 &2   +0.14 0.01 &25   +0.25 0.18\\
$[$Sc/Fe$]${\sc ii}  &7 $-$0.08 0.12 &69   +0.00 0.09 &2 $-$0.22 0.02 &25 $-$0.17 0.11\\
$[$Ti/Fe$]${\sc i}   &7   +0.18 0.10 &69   +0.16 0.14 &2   +0.03 0.08 &25   +0.08 0.11\\
$[$Ti/Fe$]${\sc ii}  &7   +0.27 0.12 &		      &2   +0.08 0.21 & 	      \\
$[$V/Fe$]${\sc i}    &7 $-$0.07 0.09 &69 $-$0.16 0.16 &2   +0.17 0.25 &21 $-$0.01 0.16\\
$[$Cr/Fe$]${\sc i}   &7   +0.06 0.09 &69 $-$0.03 0.09 &2 $-$0.10 0.03 &25 $-$0.07 0.14\\
$[$Mn/Fe$]${\sc i}   &7 $-$0.49 0.09 &		      &2   +0.01 0.14 & 	      \\
$[$Fe/H$]${\sc i}    &7 $-$1.51 0.16 &69 $-$1.56 0.19 &2 $-$0.74 0.22 &25 $-$0.61 0.36\\
$[$Fe/H$]${\sc ii}   &7 $-$1.48 0.17 &69 $-$1.52 0.16 &2 $-$0.73 0.11 &18 $-$0.69 0.39\\
$[$Co/Fe$]${\sc i}   &7 $-$0.15 0.15 &45 $-$0.08 0.06 &2 $-$0.29 0.11 &19 $-$0.20 0.09\\
$[$Ni/Fe$]${\sc i}   &7 $-$0.09 0.03 &69 $-$0.10 0.06 &2 $-$0.17 0.07 &25 $-$0.13 0.05\\
$[$Cu/Fe$]${\sc i}   &7 $-$0.61 0.18 &                &2 $-$0.55 0.07 &               \\
\hline
\end{tabular}
\label{t:meanabu}
\end{table*}

Abundances of all species in the individual stars are given in 
Table~\ref{t:proton}, Table~\ref{t:alpha}, and Table~\ref{t:fegroup}, only
available on line at CDS.

\subsection{$\alpha-$elements}

For Ti, lines of both neutral and singly ionized species were available in the 
spectral range of UVES spectra; the good ionization equilibrium we found both 
for Fe and Ti (see Table~\ref{t:meanabu}) supports the reliability of the 
adopted parameters, also for stars of the SgrN.

\begin{figure} 
\centering
\includegraphics[scale=0.54]{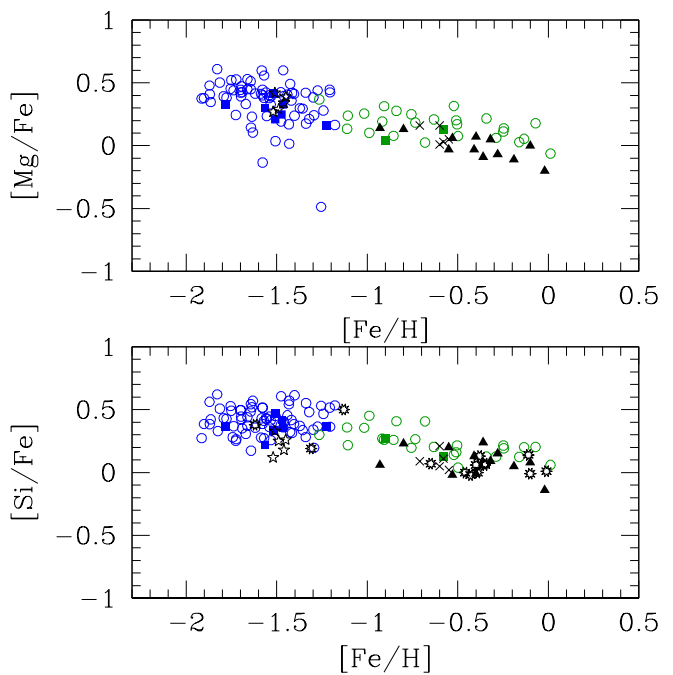}
\caption{Run of abundance ratios for the $\alpha-$elements [Mg/Fe] and 
[Si/Fe]. Symbols are as in Fig.~\ref{f:m54ona}.}
\label{f:m54mgsi}
\end{figure}

\begin{figure} 
\centering
\includegraphics[scale=0.54]{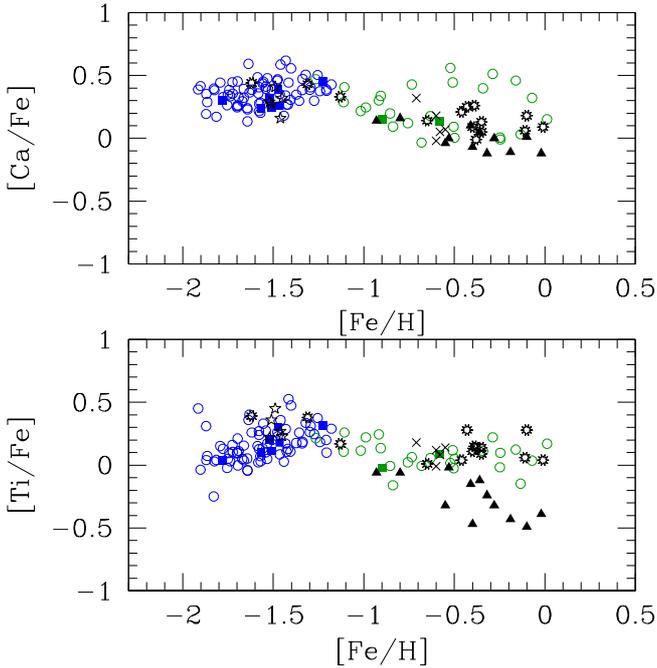}
\caption{Run of abundance ratios for the $\alpha-$elements [Ca/Fe] and 
[Ti/Fe]. Symbols are as in Fig.~\ref{f:m54ona}.}
\label{f:m54cati}
\end{figure}

The run of the $\alpha-$elements Mg, Si, Ca and Ti~{\sc i} as a function of
metallicity are shown in Fig.~\ref{f:m54mgsi} and Fig.~\ref{f:m54cati}.
We remind that some stars in M~54 appear to be quite Mg-poor, showing 
signature of extensive processing by H-burning at high temperature. For the 
Sgr galaxy, on the other hand, a rather tight relation between [Mg/Fe] and 
[Fe/H] is visible in the upper panel of Fig.~\ref{f:m54mgsi}, with maybe a 
hint of a residual offset between our values and those by S07. No Mg-poor 
stars appear in this component at any
metallicity, confirming that the stars in SgrN are not formed from gas enriched
by light elements as in the case of cluster stars.
Stars in M~54 by BWG99 and the few giants in Ter~7 from S07 are very
well mixed with our M~54 sample and with the stars in Sgr nucleus, respectively.

The [Si/Fe] ratios run almost flat as a function of the metallicity in the 
whole M~54 sample;  again, an offset seems to be present in the average
[Si/Fe] ratio for M~54 between us and BWG99 (lower panel in
Fig.~\ref{f:m54mgsi}).

We found a larger scatter in our [Ca/Fe] ratios among our SgrN stars
with respect to literature sample and, {\it viz}, a smaller scatter concerning
Ti; in particular, [Ti/Fe] ratios derived by S07 seem to be very low when
compared to abundances from our study and from SHM09. Note that
adding stars from Monaco et al. (2005) would not alleviate the discrepancy, since
they would fall half way between S07 values and ours+SHM09.
In M~54 we found that [Ca/Fe] slightly increases as [Fe/H] increases; 
this same feature is present also in the next-of-kin cluster $\omega$~Cen, well
evident in the NDC95 data, and in the recent analysis by CJ09.

In all plots of $\alpha-$elements we can pick up similar features. The
metal-poor component, represented by M~54 stars, has high levels of 
$\alpha-$elements which are typical of the Galactic metal-poor (halo)
component. 
At higher metallicities there is a smooth, steady decline in all abundance 
ratios of the Sgr field component. This pattern, offset in the high [Fe/H]
regime at lower [$\alpha$/Fe] ratios with respect to the Galactic disk
population, is present both in our data and in those of Sgr field stars by
SHM09. This is a well known feature of Sgr since the first large samples of
stars in this disrupting galaxy were gathered for abundance analysis (see
Smecker-Hane and McWilliam 2002, 2009;  S07, and references
therein). 

Elements deriving by $\alpha-$capture processes are produced almost exclusively 
in massive,
fast evolving stars ending their life as core-collapse SNe (e.g. Wheeler et al.
1989 and references therein).  
The elemental ratio [$\alpha$/Fe], and in particular the knee point as a
function of metallicity, is driven by the {\it clock} of stellar evolution,
since the major producers of iron (and iron-peak elements) are the smaller mass
stars exploding as type Ia SNe after a time delay of about 0.5-3 Gyr with
respect to the original episode of star formation (Tinsley 1979, Matteucci \&
Fran\c cois 1989).
The high and almost constant level of $\alpha-$elements in M~54 shows that
stars in this cluster formed before the typical $e-$folding time for SN Ia
contributing their ejecta to the gas pool. The complete lack of stars more metal
poor than [Fe/H]$\sim -2$ also suggests that the burst of star formation
originating M~54 probably occurred in a system already enriched from ejecta of
massive stars.

From what it is observed in the Sgr galaxy (Fig.~\ref{f:m54mgsi} and 
Fig.~\ref{f:m54cati}) it is not easy to detect a turn down point in every plot.
However, if we consider M~54 itself as a  representative sample of the
metal-poor component of the Sgr complex\footnote{A recent extensive study by
Chou et al. (2009) also reports that farther along the Sgr stream an increasing 
number of metal-poor stars is observed to show abundances typical of Galactic
halo stars and similar to those of the metal-poor cluster M~54 (see also Monaco
et al. 2007).}  we can conclude that a knee can be confidently located at
metallicities in the range [Fe/H]$=-1.4 \div -1.3$. By averaging Si, Ca
and Ti - silicon being only slightly involved in proton-capture reactions -
the position of the knee seems to be confirmed at [Fe/H]$\sim -1.3$ dex.
This feature confirms once more that Sgr was likely subject to a 
lower star formation rate in its evolution than the
Milky Way, whose knee occurs at a metallicity of $\sim -1$.

\subsection{Fe-peak elements}

\begin{figure} 
\centering
\includegraphics[scale=0.54]{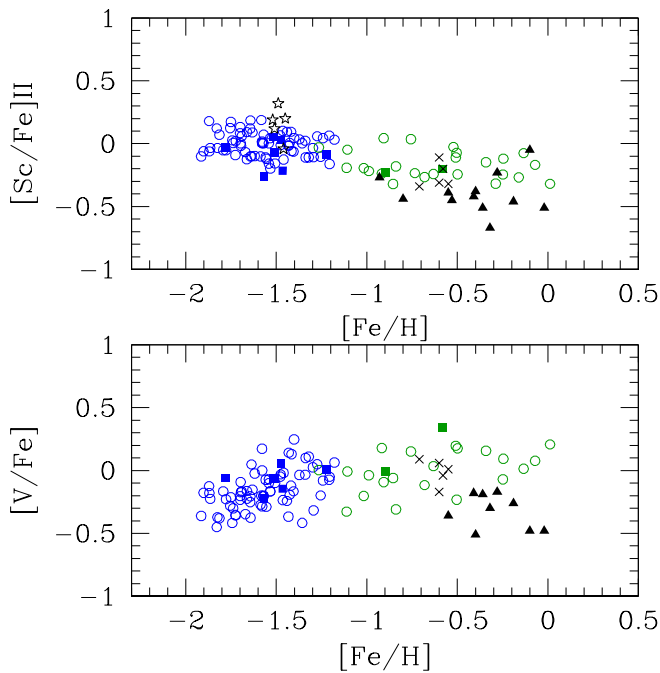}
\caption{Run of abundance ratios for Sc~{\sc ii} and V~{\sc i}. Symbols are as in
Fig.~\ref{f:m54ona}.}
\label{f:m54scv}
\end{figure}

\begin{figure} 
\centering
\includegraphics[scale=0.54]{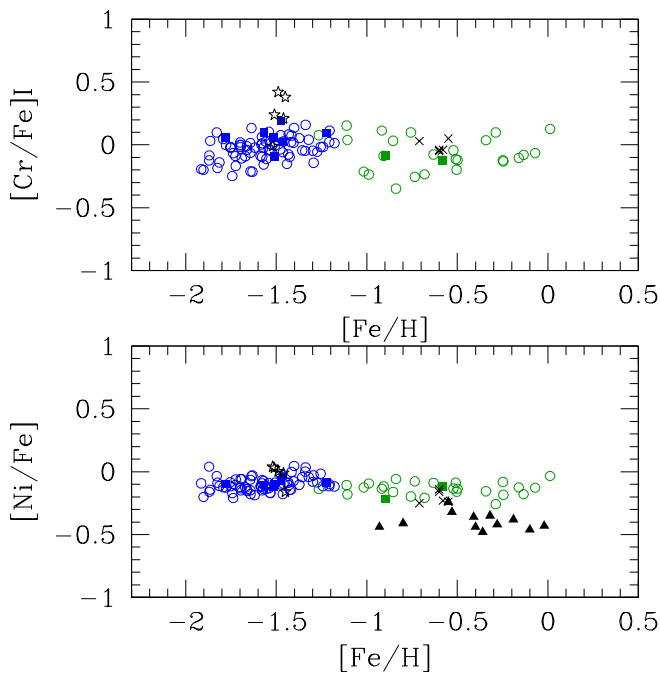}
\caption{Run of abundance ratios for Cr~{\sc i} and Ni~{\sc i}.}
\label{f:m54crni}
\end{figure}

Abundances of Sc~{\sc ii} , V~{\sc i}, Cr~{\sc i} and Ni~{\sc i} 
are displayed as a function of the metallicity in
Fig.~\ref{f:m54scv}, Fig.~\ref{f:m54crni} 
for stars analyzed in M~54 and SgrN.

On average (see Table~\ref{t:meanabu}) these elements closely track Fe in M~54,
with a flat distribution around a solar [El/Fe] ratio for both the metal-rich
and metal-poor components (maybe apart from V {\sc i}). On the other hand, all
elements  belonging the Fe-group show a small deficiency with respect to the
solar ratio in  stars of SgrN (again, with the exception of V). The offets 
observed in the run of Fe-group elements with respect to previous analysis are
easily accounted for by the different scales of adopted atmospheric and atomic 
parameters.

\begin{figure} 
\centering
\includegraphics[scale=0.40]{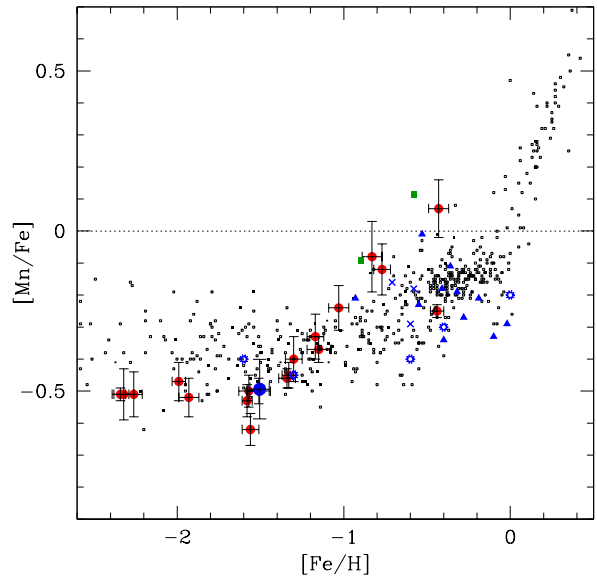}
\caption{[Mn/Fe] as a function of metallicity for field stars (tiny squares)
from several studies (see text); the average values for M~54 (this study, large
blue filled circle) and 19 GCs of this project (Gratton et al. 2006, Carretta et
al. 2007b, Carretta et al, in preparation) are indicated by red filled circles, 
with 1$\sigma$ 
error bars. Green filled squares are the [Mn/Fe] ratios for 2 stars of Sgr with
UVES spectra in the present study. Crosses and filled triangles are stars in
Terzan~7 and in the Sgr main body from Sbordone et al. (2007). Finally, "star
explosion" symbols are for stars in Sgr from McWilliam et al. (2003).}
\label{f:mnfe2}
\end{figure}

The run of [Mn/Fe] as a function of the metallicity (Fig.~\ref{f:mnfe2}) is 
important because Mn production is related to the
available neutron excess (see Arnett 1971). By comparing Mn abundances in
environments with different metallicity, McMilliam et al. (2003)
concluded that Mn yields were metallicity-dependent both in type Ia SNe and in
core-collapse SNe. As a consequence, the deficiency of Mn observed in metal-rich
stars of Sgr can be interpreted as an evidence of a significant contribution 
from metal-poor SNe Ia, even at high metallicities. Therefore, the yields
from these metal-poor SNe, incorporated in the gas from where the metal-rich Sgr
component formed, could explain the underabundance in [Mn/Fe] observed in Sgr
stars at high metal abundance, when compared with Galactic field stars at the
same metallicity.

\begin{figure} 
\centering
\includegraphics[scale=0.40]{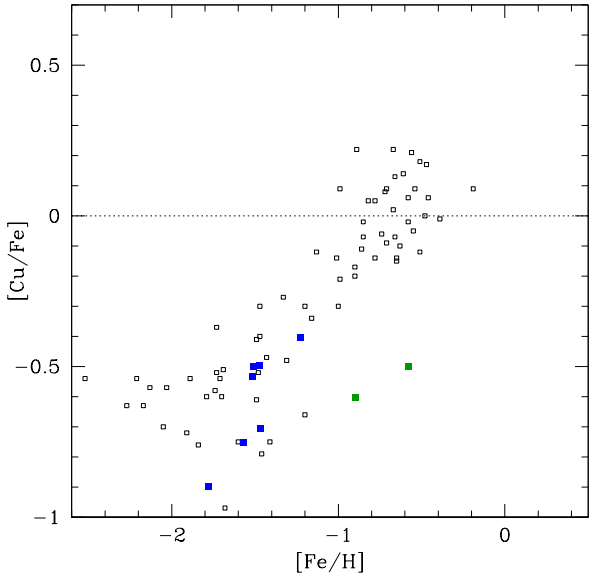}
\caption{The ratio [Cu/Fe] as a function of metallicity for field stars 
(tiny squares) of the Galactic halo and disk from Mishenina et al. (2002) and
for the stars of M~54 and SgrN with UVES spectra (blue and green large filled 
squares, respectively) from the present study.}
\label{f:cum54}
\end{figure}

The same conclusion is obtained from the abundance ratios [Cu/Fe] (see McWilliam
and Smecker-Hane 2005b). The large deficiency in Cu for metal-rich Sgr stars,
is visible also in our data (albeit quite sparse, see Fig.~\ref{f:cum54}). 
This underabundance, with
respect to field stars in the Galaxy, can be explained as due to a substantial
contribution from metal-poor SN Ia, since Cu is almost entirely produced by 
SN II, whose Cu yields are metallicity dependent (Woosley and Weaver 1995).
At the high metallicity regime sampled by the SgrN component the only way 
to reproduce such a deficiency is then to invoke an extra contribution from
metal-poor type Ia SNe to the gas pool from which these stars formed.
Here we only note that our stars in M~54 are well intermingled with those
attributed by McWilliam and Smecker-Hane (2005b, their Fig.1) to the metal-poor
component of the Sgr nucleus, stressing once again the difficulty of
disentangling cluster and nucleus at these low metallicities.

The analysis of elements produced by neutron-capture processes in stars of M~54
and SgrN will be presented in a separate paper.

\section{Testing the scenario for GC formation}

\subsection{General considerations}

In Carretta et al. (2010a) we noticed that the very same definition of
{\it globular cluster} commonly used is not entirely unambiguous. There are old
open clusters whose masses and sometime ages may exceed those of populous
stellar cluster in the Magellanic Clouds and that lie on the extension of the
locus populated by Galactic GCs in the $M_V$ vs age plane.  To draw an objective
distinction we then dubbed as {\it bone fide} globular clusters those where a
Na-O anticorrelation is found. Already Carretta (2006) speculated that this 
feature is so universal among GCs that it should be related to their origin. 
We then expanded this view to a comprehensive, albeit still qualitative, 
scenario for the formation of GCs as distinct from other systems of similar 
mass (like dwarf spheroidal galaxies, see Carretta et al. 2010a), for which 
the Na-O anticorrelation is not observed. Indeed, whenever data are available 
for both GCs and field stars in the same galaxy (Fornax - Letarte et al. 2006, 
Shetrone et al. 2003, Sgr - this study) {\it only} stars in cluster do actually 
participate to the Na-O anticorrelation.

The main discriminating factor between GCs and dwarf dSphs is quite evident when
comparing the two massive clusters examined in this paper with typical low mass dSphs: 
the two GCs lie at least 3 times (M~54, and much more $\omega$ Cen) 
closer to the central potential well of the Milky Way than any of the dSphs,
except for Sgr itself.
This obvious observation spurred us to outline in Carretta et al. (2010a) a
model where the ancestors of both dSphs and GCs started as cosmological
fragments, but those that ultimately evolved into currently observed GCs started
their evolution much closer to the central region of our Galaxy.

Very briefly, the model sketched predicts that strong interactions with the main
Galaxy trigger the transformation of some gas in stars in the first phases. 
This event forms a $precursor$ whose core-collapse SNe enrich the
system in metals, triggering a burst of star formation. At the end, an expanding
association free of the primordial gas, swept away by energetic winds by SN II
and massive  stars, is left. After a while, the low velocity winds from evolving
fast  rotating massive stars or intermediate-mass AGB stars replenish (or a
cooling flow re-collects in) the  central regions a new gas reservoir. It is
only in this phase that a second generation of stars may be born as a very compact
central cluster, which is what we currently are observing as a GC, together
with the primordial stars.

The advantage of this scenario is to include a coherent picture the different
stellar generations in GCs (Gratton et al. 2001, Carretta et al. 2009a) within a
global view of the Galaxy formation. This approach is not  new, starting with
works by Searle and Zinn (1978). Several supports can be found in the scientific
literature: dynamical  interactions are known to be a strong trigger of star
formation in galactic satellites, as suggested both by observations (Zaritsky
and Harris 2004) and theoretical cosmological simulations (e.g. Kravtsov et al.
2004).

Moreover, we provided in Carretta et al. (2010a) several arguments suggesting
that the present
GCs started their evolution as objects whose baryonic masses were at least
20 times larger than the currently observed masses, and successively lost a large
fraction of their primordial mass (as also found by theoretical modeling, see e.g.
Baumgardt et al. 2008, D'Ercole et al. 2008). These arguments (including
metallicity, chemistry, kinematic and density distribution) will not be
repeated; here we limit to highlight two other convincing evidences that
are appropriate when discussing cases such as M~54 and $\omega$ Cen, at the 
very high mass end of the mass distribution of GCs.

The first evidence comes from the comparison between nuclear star clusters, 
residing at the centre of dwarf galaxies, and GCs with extended
hot HBs recently made by Georgiev et al. (2009). In their Figure 4b they show
that in the mass-size plane (as represented by $M_V$ vs half mass radius $r_h$)
massive Galactic GCs occupy the same region populated by nuclear
star clusters, but GCs like M~54 and $\omega$ Cen are offset toward larger
sizes. They propose that this finding can be explained by the expansion of
the tidal radius, as well as of the $r_h$, that follows the accretion of the
fragments into the halo of a much more massive galaxy. This evidence is quite
appealing, since it nicely fits in several ways our outlined scenario, where
all GCs originate in much larger fragments: M~54 and $\omega$~Cen illustrate 
in two snapshots the likely mechanism of loss of the surrounding host galaxy, 
already completed for $\omega$~Cen and still ongoing in the case of M~54. 

The second point may be made repeating for M~54 an argument originally made 
by Suntzeff and Kraft (1996) for a closed box chemical evolution model of 
$\omega$~Cen. From studies in the solar neighbourhood we know that the yield 
per stellar generation is about 0.01 (Tinsley 1979); this means that some 1\% 
of the available gas is transformed into metals in a single stellar generation. 
On the other hand, we found that metallicities for stars in M~54 range from 
[Fe/H]$=-1.6$ to [Fe/H]$=-1.3$ dex, i.e. from Z$\sim 0.0005$ to Z$\sim 0.001$, 
with a change of $\Delta$Z=0.0005. This is $1/20$ of the yield of a stellar 
generation. We are then left with only two options: (i) to selectively lose 
the metal-rich gas component (lowering the effective yield per stellar generation) 
or (ii) to lose 95\% of the gas before transforming into metals. In the second 
case, and considering the narrow mass range of the polluters required to
produce the O-Na anticorrelation, we conclude that the precursors had a baryonic 
mass some 200 times larger than M~54 and $\omega$~Cen. This is comparable to
the current baryonic mass of the Sgr dwarf galaxy (Majewski et al. 2003; 
Niederste-Ostholt et al. 2010).

\subsection{Extension of the model to the high mass range}

As sketched in Carretta et al. (2010b), the results we obtained for M~54 (and 
by extension for $\omega$ Cen) might be explained by a very simple extension 
of the scenario for cluster formation we considered insofar (Carretta et al. 
2010a). The main modification is to switch from a single cooling flow to 
multiple, nearly simultaneous cooling flows, or episodes of gas gathering. 
We think this is not unlikely in the most massive objects.

To clarify what we are saying, let us first recall the main observational 
facts:
\begin{itemize}
\item[1.] M~54 has a range in metal abundances of a few tenths of a dex.
\item[2.] The Na-O anticorrelation can be observed at all metallicities.
\item[3.] However, the anticorrelation is more extended at higher metallicities,
where we observe a large incidence of extremely O-poor, likely very
He-rich stars (that is, the very He-rich population is more metal rich
than the average of the He-normal population).
\end{itemize}

We note that fact 1. can be easily explained if the region where the primordial
population forms is extended, so much that Fe abundances cannot be well
homogenized (likely by turbulence). On the whole, this might be expected
for large enough star forming regions. Fact 2. requires multiple cooling
flows, with different regions of the primordial population producing
individual clusters (that later merged). Again, this might be expected if
the star forming region is large enough. However, these two facts may also
be explained by two different and separate cluster formation episodes. The
real intriguing fact is the third one, which is also evident in the case of $\omega$
Cen; since it occurred in both the most massive GCs of our galaxy, it is quite
unlikely to be the result of a random sequence of events.

In our interpretative scenario, the explanation of fact 3. requires an
appropriate timing. In fact, we expect the extension of the
anticorrelation to be determined by the mass of the polluters. 
In the case of intermediate-mass AGB stars, large mass
polluters (6-8 $M_\odot$, lifetimes $\sim 40-70$ Myr, using the isochrones by 
Marigo et al. 2008) produce an extended anticorrelation, with very low O 
abundances, and large production of He. Smaller mass polluters (5-6 $M_\odot$, 
lifetime $\sim 70-100$ Myr) produce a much less extended anticorrelation, with 
minimal effects on He. 
Hence, in order to explain the observations we need that the metal-poor
second generation is formed by the ejecta of intermediate mass AGB stars,
while the metal-rich one is formed ($only$) by the ejecta of the very
massive ones. The easiest way to avoid contribution by massive polluters
is to delay the cooling flow (or gas replenishing): so we need to delay 
the flowing of the gas for the
metal-poor component by about 10-30 Myr. This can easily be obtained if we
assume that the metal-rich primordial component formed $\sim 10-30$ Myr $later$ 
than the
metal-poor primordial component. In this case, while the 6-8 $M_\odot$ stars 
of the metal-poor
component are in their AGB phase, massive stars of the metal-rich component are
still exploding as core-collapse SNe. This might cause a considerable
input of energy in the system, preventing the formation of the cooling
flow, anyway of gas collection, until the rate of SN explosion becomes 
small enough. Then, a quiet
phase follows, lasting some tens Myr. In this phase, cooling flow
formation is possible for both the metal-poor component (with AGB
polluters of 5-6 $M_\odot$), and for the metal-rich one (with AGB polluters 
of 6-8 $M_\odot$). The second generation stars form nearly simultaneously in 
the metal-poor and metal-rich cooling flows: finally, the SN explosion of
stars formed in these cooling flows (or the onset of type Ia SNe) stops
this later phase of star formation.

This scenario is actually very similar to what we had already considered for the
formation of small clusters (in groups, see Carretta et al. 2010a). The only
basic difference is that here we assume that also the adjacent star forming
regions produce a cooling flow. Hence, the main difference is that in this
working  hypothesis the star formation continues at a high rate for a prolonged
period (although shifting toward other regions of the global star forming area),
while in the case of the small mass clusters the star formation declines
earlier, though not so fast that the residual star formation and then late SNe
have not an impact on the formation of the cooling flow.

Hidden in this scenario is the hypothesis that metallicity raises with time
within the star forming region, i.e. that those areas where star formation
occurred later are also more metal-rich, which is reasonable but not
entirely obvious.

Of course, this scenario needs verifications (e.g. for what concerns the 
neutron-capture elements) and it is for the moment quite speculative. However, 
we think it is on the whole a reasonable extension to larger masses of what we 
have considered insofar for more typical GCs.

To better understand the mechanism we put forward, it is necessary to have
a deeper look to what it is known about star formation, which is still
far from being properly understood.
However, various authors (see the review by McKee and Ostriker 2007 and
references therein) suggest that star formation within a Giant
Molecular Cloud (GMC) occurs on time-scales comparable to a few times
(1-3) the crossing time, which is roughly t$_{cr} \sim 10 M_6^{1/4}$ Myr, 
where M$_6$
is the mass of the cloud in units of 10$^6 M_\odot$. Using this relation, we
obtain that the crossing time is:
\begin{itemize}
\item 5.6 Myr for 10$^5 M_\odot$,
\item 10 Myr for 10$^6 M_\odot$,
\item 18 Myr for 10$^7 M_\odot$,
\item 32 Myr for 10$^8 M_\odot$.
\end{itemize}

Keeping in mind that the final cluster is a few percent of the original
cloud, a timescale of 30 Myr for the formation of the primordial
population of a cluster like M~54 and $\omega$ Cen (currently a few 10$^6$ yr)
seems fully reasonable. Clusters smaller in size by an order of magnitude
(that is, typical clusters with M$_V \sim- 7.8$) should have required shorter
time ($\sim 10-20$ Myr).

On the other hand, Galactic GMCs seem to have an upper limit to their
mass, of a few 10$^6$ $M_\odot$; larger masses are usually called Giant Molecular
Association, because they fragment in smaller pieces. It is not clear if
this depends on metallicity or environment. This is likely the reason why GCs
do not form anymore in the  Milky Way, given the low efficiency with which
GMCs can be transformed in clusters. Also, fragmentation of the larger
molecular clouds may lead to more complicate geometries, and then to
smaller efficiencies in the cloud-to-cluster transformation.

While these arguments are purely qualitative, they do not appear
inconsistent with the scenario we are proposing. Of course, fine tuning is
required to properly match observations, but at least the order of
magnitudes for the relevant quantities seem to be correct.

\section{Summary}

In this paper we present the most extensive survey, up to date, of the chemical
composition for red giant stars in the GC M~54, which lies into 
the nucleus of the Sgr dwarf
galaxy. We derived detailed abundances of elements involved in proton-capture 
reactions of H-burning at high temperature (O, Na, Mg, Al, Si), of
$\alpha-$elements, and of Fe-peak elements for 76 RGB stars in M~54 from FLAMES
spectra. The abundances of the same species are also homogeneously obtained for
a sample of 27 stars of the nuclear component of Sgr.

For the first time the metallicity spread within this cluster is obtained and
well constrained from high resolution spectra. The iron spread found is
intermediate between that observed in smaller, normal Galactic GCs and the
dispersion in metallicity associated to the few dwarf galaxies investigated so
far. This finding supports the idea that both M~54 and $\omega$ Cen can be
identified as nuclear star clusters, once (or still, in the first case) in the
nucleus of dwarf galaxies. In Carretta et al. (2010b) we already showed that the
most metal-rich tail of the metallicity distribution is given by SgrN stars in
the case of M~54 and restricted to stars on the anomalous RGB in the case of
$\omega$ Cen.

Comparing our results 
to literature data for $\omega$ Cen, we find that these two massive
clusters are similar in several characteristics, the most notable being the
presence of multiple populations, well evident both from spectroscopy and
photometry. 
In particular, the Na-O and the Mg-Al anticorrelations among proton-capture
elements is present for both the metal-poor and the metal-rich stellar component
in each cluster, being more extended for the second one. 

We know that the Na-O anticorrelation is present in all GCs, that 
is related to the
existence of multiple stellar generations so that it may be considered the main
signature of a {\it bona fide} globular cluster, as distinct from the star
formation in other environment (Carretta et al. 2010a). As a consequence, the
absence of the anticorrelation in the very metal-rich component at near solar
metallicities both in M~54 and $\omega$ Cen led us to conclude that this
population was formed with a different channel, probably being composed by the
host galaxy (still observable in the case of M~54).

The presence and different extension of the Na-O anticorrelation (also supported
by the Mg-Al data) for the metal-poor and moderately metal-rich $cluster$
components suggest a simple extension of the scenario for the formation
of GCs illustrated in Carretta et al. (2010a). 
We need (i) that the primordial population of the metal-rich component is 
enriched in metals with respect to the metal-poor component, and (ii) that the 
mass range of the polluters acting to form the second generation is different
for the two components (with more massive polluters at work for the metal-rich
one, where the anticorrelations are more extended).

We showed that the simplest way to explain the observed features in M~54 (and
for analogy in $\omega$ Cen) is to assume that the primordial population of the
moderately metal-rich component formed with a small delay (as much as 10-30 Myr)
after the corresponding population of the metal-poor component, whereas the
second generation was formed more or less simultaneously in both components.

\begin{acknowledgements}
This publication makes use of data products from the Two Micron All Sky
Survey, which is a joint project of the University of Massachusetts and the
Infrared Processing and Analysis Center/California Institute of Technology,
funded by the National Aeronautics and Space Administration and the National
Science Foundation. We warmly thank Andy McWilliam for sending us his manuscript
and data on Sgr in advance of publication. We thank Jennifer Sobeck for sharing
the Cu linelist in advance of publication. This work was partially funded by the
Italian MIUR under PRIN 2003, 2007 and by the grant INAF 2005 ''Experimenting
nucleosynthesis in clean environments''.  S.L. is grateful to the DFG cluster of
excellence ''Origin and Structure of the Universe'' for partial support.  M.B.
acknowledges the financial support of INAF through the PRIN-2007 grant CRA
1.06.10.04.  This research has made use of the SIMBAD database, operated at CDS,
Strasbourg, France and of NASA's Astrophysical Data System.
\end{acknowledgements}

\begin{appendix}

\section{Error estimates}
We refer the reader to the analogous Appendices in Carretta et al. (2009a,b) for
a detailed discussion of the procedure adopted for error estimates. 
In the following we only provide the main tables of sensitivities of abundance
ratios to the adopted errors in the atmospheric parameters and $EW$s and the
final estimates of internal and systematic errors for all species analyzed from
UVES and GIRAFFE spectra of stars in M~54 and in SgrN.

\paragraph{Sensitivities of abundance ratios to atmospheric parameters.}  The
sensitivities of derived  abundances on the adopted atmospheric parameters were
obtained by repeating our abundance analysis by changing only one atmospheric
parameter each time for $all$ stars in M~54 and in the nucleus of Sgr,
separately, and taking the average value of the slope change vs abundance.
This exercise was done for both UVES and GIRAFFE spectra.

The amount of the variations in the atmospheric parameters is shown in the
first line of the headers in Table~\ref{t:sensitivityu15},
Table~\ref{t:sensitivitym15}, and Table~\ref{t:sensitivitym14}, 
whereas the resulting 
response in abundance changes of all elements (the sensitivities) are shown 
in columns from 3 to 6 of these tables, for abundances in M~54 from UVES and
GIRAFFE spectra and for stars in the Sgr nucleus from GIRAFFE, respectively.

\begin{table*}
\centering
\caption[]{Sensitivities of abundance ratios to variations in the atmospheric
parameters and to errors in the equivalent widths, and errors in abundances for
stars in M~54 observed with UVES}
\begin{tabular}{lrrrrrrrr}
\hline
Element     & Average  & T$_{\rm eff}$ & $\log g$ & [A/H]   & $v_t$    & EWs     & Total   & Total      \\
            & n. lines &      (K)      &  (dex)   & (dex)   &kms$^{-1}$& (dex)   &Internal & Systematic \\
\hline        
Variation&             &  50           &   0.20   &  0.10   &  0.10    &         &         &            \\
Internal &             &   5           &   0.04   &  0.06   &  0.05    & 0.060   &         &            \\
Systematic&            &  38           &   0.06   &  0.05   &  0.02    &         &         &            \\
\hline
$[$Fe/H$]${\sc  i}& 70 &   +0.056      & $-$0.045 &$-$0.001 & $-$0.026 & +0.007  &0.018    &0.050      \\
$[$Fe/H$]${\sc ii}&  8 &   +0.077      &   +0.095 &  +0.031 & $-$0.011 & +0.021  &0.035    &0.090      \\
$[$O/Fe$]${\sc  i}&  2 & $-$0.042      &   +0.070 &  +0.034 &   +0.025 & +0.042  &0.051    &0.182      \\
$[$Na/Fe$]${\sc i}&  3 & $-$0.007      & $-$0.055 &$-$0.034 &   +0.015 & +0.035  &0.042    &0.149      \\
$[$Mg/Fe$]${\sc i}&  3 & $-$0.020      & $-$0.017 &$-$0.006 &   +0.016 & +0.035  &0.036    &0.037      \\
$[$Al/Fe$]${\sc i}&  2 & $-$0.011      & $-$0.023 &$-$0.009 &   +0.015 & +0.042  &0.044    &0.255      \\
$[$Si/Fe$]${\sc i}&  8 & $-$0.065      &   +0.020 &  +0.010 &   +0.020 & +0.021  &0.025    &0.058      \\
$[$Ca/Fe$]${\sc i}& 17 &   +0.014      & $-$0.025 &$-$0.018 & $-$0.014 & +0.015  &0.020    &0.033      \\
$[$Sc/Fe$]${\sc i}&  1 &   +0.036      & $-$0.028 &$-$0.014 &   +0.021 & +0.060  &0.062    &0.059      \\
$[$Sc/Fe$]${\sc ii}& 7 & $-$0.084      & $-$0.020 &$-$0.003 & $-$0.007 & +0.023  &0.025    &0.078      \\
$[$Ti/Fe$]${\sc i}&  9 &   +0.044      & $-$0.020 &$-$0.017 &   +0.004 & +0.020  &0.023    &0.052      \\
$[$Ti/Fe$]${\sc ii}& 1 & $-$0.093      & $-$0.013 &  +0.000 &   +0.007 & +0.060  &0.061    &0.083      \\
$[$V/Fe$]${\sc i} &  8 & $-$0.029      & $-$0.103 &$-$0.113 & $-$0.090 & +0.021  &0.085    &0.054      \\
$[$Cr/Fe$]${\sc i}&  2 &   +0.023      & $-$0.022 &$-$0.013 &   +0.016 & +0.042  &0.044    &0.037      \\
$[$Mn/Fe$]${\sc i}&  3 &   +0.009      & $-$0.017 &$-$0.009 &   +0.000 & +0.035  &0.035    &0.036      \\
$[$Co/Fe$]${\sc i}&  4 & $-$0.018      & $-$0.004 &  +0.001 &   +0.020 & +0.030  &0.032    &0.058      \\
$[$Ni/Fe$]${\sc i}& 20 & $-$0.023      &   +0.012 &  +0.005 &   +0.004 & +0.013  &0.014    &0.020      \\
$[$Cu/Fe$]${\sc i}&  1 & $-$0.031      &   +0.012 &  +0.001 & $-$0.007 & +0.013  &0.060    &0.070      \\
\hline
\end{tabular}
\label{t:sensitivityu15}
\end{table*}

\begin{table*}
\centering
\caption[]{Sensitivities of abundance ratios to variations in the atmospheric
parameters and to errors in the equivalent widths, and errors in abundances for
stars in M~54 observed with GIRAFFE}
\begin{tabular}{lrrrrrrrr}
\hline
Element     & Average  & T$_{\rm eff}$ & $\log g$ & [A/H]   & $v_t$    & EWs     & Total   & Total      \\
            & n. lines &      (K)      &  (dex)   & (dex)   &kms$^{-1}$& (dex)   &Internal & Systematic \\
\hline        
Variation&             &  50           &   0.20   &  0.10   &  0.10    &         &         &            \\
Internal &             &   5           &   0.04   &  0.05   &  0.07    & 0.077   &         &            \\
Systematic&            &  38           &   0.06   &  0.04   &  0.01    &         &         &            \\
\hline
$[$Fe/H$]${\sc  i}& 38 &   +0.055      &   +0.006 &$-$0.004 & $-$0.031 & +0.012  &0.026    &0.042      \\
$[$Fe/H$]${\sc ii}&  3 & $-$0.044      &   +0.092 &  +0.027 & $-$0.008 & +0.044  &0.050    &0.047      \\
$[$O/Fe$]${\sc  i}&  1 & $-$0.041      &   +0.074 &  +0.036 &   +0.029 & +0.077  &0.083    &0.055      \\
$[$Na/Fe$]${\sc i}&  3 & $-$0.009      & $-$0.051 &$-$0.029 &   +0.022 & +0.044  &0.050    &0.041      \\
$[$Mg/Fe$]${\sc i}&  2 & $-$0.014      & $-$0.015 &$-$0.005 &   +0.017 & +0.054  &0.056    &0.024      \\
$[$Si/Fe$]${\sc i}&  7 & $-$0.059      &   +0.024 &  +0.012 &   +0.025 & +0.029  &0.035    &0.047      \\
$[$Ca/Fe$]${\sc i}&  7 &   +0.017      & $-$0.029 &$-$0.013 & $-$0.013 & +0.029  &0.032    &0.020      \\
$[$Sc/Fe$]${\sc ii}& 6 & $-$0.061      &   +0.068 &  +0.031 &   +0.010 & +0.031  &0.038    &0.052      \\
$[$Ti/Fe$]${\sc i}&  5 &   +0.039      & $-$0.016 &$-$0.014 &   +0.008 & +0.034  &0.036    &0.034      \\
$[$V/Fe$]${\sc i} &  6 &   +0.056      & $-$0.015 &$-$0.015 &   +0.013 & +0.031  &0.034    &0.047      \\
$[$Cr/Fe$]${\sc i}&  4 &   +0.020      & $-$0.018 &$-$0.009 &   +0.024 & +0.039  &0.042    &0.020      \\
$[$Co/Fe$]${\sc i}&  1 &   +0.004      &   +0.009 &  +0.007 &   +0.025 & +0.077  &0.079    &0.010      \\
$[$Ni/Fe$]${\sc i}&  8 & $-$0.014      &   +0.014 &  +0.009 &   +0.021 & +0.027  &0.031    &0.013      \\
\hline
\end{tabular}
\label{t:sensitivitym15}
\end{table*}

\begin{table*}
\centering
\caption[]{Sensitivities of abundance ratios to variations in the atmospheric
parameters and to errors in the equivalent widths, and errors in abundances for
stars in SgrN observed with GIRAFFE}
\begin{tabular}{lrrrrrrrr}
\hline
Element     & Average  & T$_{\rm eff}$ & $\log g$ & [A/H]   & $v_t$    & EWs     & Total   & Total      \\
            & n. lines &      (K)      &  (dex)   & (dex)   &kms$^{-1}$& (dex)   &Internal & Systematic \\
\hline        
Variation&             &  50           &   0.20   &  0.10   &  0.10    &         &         &            \\
Internal &             &  18           &   0.08   &  0.01   &  0.18    & 0.053   &         &            \\
Systematic&            &  53           &   0.09   &  0.02   &  0.04    &         &         &            \\
\hline
$[$Fe/H$]${\sc  i}& 26 &   +0.000      &   +0.040 &  +0.018 & $-$0.049 & +0.010  &0.090    &0.027      \\
$[$Fe/H$]${\sc ii}&  2 & $-$0.086      &   +0.120 &  +0.043 & $-$0.017 & +0.037  &0.076    &0.136      \\
$[$O/Fe$]${\sc  i}&  2 &   +0.014      &   +0.040 &  +0.020 &   +0.043 & +0.037  &0.088    &0.049      \\
$[$Na/Fe$]${\sc i}&  3 &   +0.041      & $-$0.070 &  +0.008 &   +0.019 & +0.031  &0.056    &0.072      \\
$[$Mg/Fe$]${\sc i}&  2 &   +0.001      & $-$0.018 &$-$0.007 &   +0.018 & +0.037  &0.050    &0.023      \\
$[$Si/Fe$]${\sc i}&  6 & $-$0.043      &   +0.017 &  +0.001 &   +0.036 & +0.022  &0.070    &0.053      \\
$[$Ca/Fe$]${\sc i}&  4 &   +0.063      & $-$0.066 &$-$0.016 & $-$0.018 & +0.027  &0.054    &0.081      \\
$[$Sc/Fe$]${\sc ii}& 4 & $-$0.012      &   +0.042 &  +0.015 &   +0.004 & +0.027  &0.033    &0.031      \\
$[$Ti/Fe$]${\sc i}&  3 &   +0.080      & $-$0.033 &$-$0.018 & $-$0.025 & +0.031  &0.063    &0.089      \\
$[$V/Fe$]${\sc i} &  4 &   +0.093      & $-$0.027 &$-$0.016 & $-$0.031 & +0.027  &0.071    &0.107      \\
$[$Cr/Fe$]${\sc i}&  5 &   +0.062      & $-$0.028 &$-$0.016 &   +0.006 & +0.024  &0.036    &0.073      \\
$[$Co/Fe$]${\sc i}&  1 &   +0.014      &   +0.014 &  +0.005 &   +0.014 & +0.053  &0.059    &0.026      \\
$[$Ni/Fe$]${\sc i}& 10 & $-$0.012      &   +0.013 &  +0.003 &   +0.027 & +0.017  &0.052    &0.020      \\
\hline
\end{tabular}
\label{t:sensitivitym14}
\end{table*}

\paragraph{Errors in atmospheric parameters} A detailed discussion of how they 
can be estimated is given in Carretta et al. (2007c); in Sect. 4 of the present
paper we describe the changes in the estimates required to take into account
the intrinsic spread in iron observed in M~54.

The other difference is that for the Sgr field (nucleus) stars, errors in
temperature were derived directly from the calibration by Alonso et al. (1999),
since we used the relation between $T_{\rm eff}(V-K)$
and the $K$ magnitudes only for the cluster M~54.

Adopted errors in atmospheric parameters are listed in the second (internal
errors) and third line (systematic errors) of the headers in 
Table~\ref{t:sensitivityu15}, Table~\ref{t:sensitivitym15}, and 
Table~\ref{t:sensitivitym14}.
These errors, multiplied for the sensitivities of abundances to variations in the
individual parameters, and summed in quadrature, provide the total internal and
systematic errors listed in the last two columns of these tables.

\end{appendix}

\end{document}